\documentclass[twocolumn,aps,pra]{revtex4-2}

\usepackage{amsmath}
\usepackage{graphicx}
\usepackage{float} 
\usepackage{subfigure} 
\usepackage[colorlinks, linkcolor=blue, anchorcolor=blue, citecolor=blue, urlcolor=blue]{hyperref}

\begin{document}

\title{Interference of Two-Dimensional Bose-Einstein Condensates in Micro-Gravity}

\author{Tie-Fu Zhang$^{1,2}$}

\author{Hao Zhu$^{1,2}$}

\author{Wen-Kai Bai$^{3}$}

\author{Kai Liu$^{1,2}$}

\author{Yi-Hui Xing$^{1,2}$}

\author{Wu-Ming Liu$^{1,2,4}$}
\email{wliu@iphy.ac.cn}

\affiliation{$^1$Beijing National Laboratory for Condensed Matter Physics, Institute of Physics, Chinese Academy of Sciences, No.8, 3rd South Street, Zhongguancun, Haidian District, Beijing, 100190, China \\ $^2$School of Physical Sciences, University of Chinese Academy of Sciences, No.19(A) Yuquan Road, Shijingshan District, Beijing, 100049, China \\$^3$Shaanxi Key Laboratory for Theoretical Physics Frontiers, Institute of Modern Physics, Northwest University, Xi'an 710127, China \\$^4$Songshan Lake Materials Laboratory, Dongguan, Guangdong 523808, China }

\begin{abstract} 
We investigate the interference of two-dimensional Bose-Einstein condensates in micro-gravity, which influenced by the interaction strength, initial momentum, gravitational potential and phase difference. We demonstrate that the gravitational potential from the Earth can change the density distribution and phase distribution of the condensate's wave function. As time evolves, a portion of the gravitational potential energy of the microscopic particles can be converted into kinetic energy, which changes the motion of the microscopic particles, and leads to the varying of the density and phase distribution of the wave function. 
Nevertheless, the influences of the Earth's gravity on the wave function can be eliminated by the micro-gravity environment, which confirmed by many micro-gravity cold atom experiments. 
Our results present the influences of gravity and other parameters on interference of Bose-Einstein condensates, which help us to reveal the intrinsic natures of the related theoretical predictions and experimental phenomena. Furthermore, our work builds a bridge between the related physical phenomena and our physical intuition about the Bose-Einstein condensates in micro-gravity environment. 
\end{abstract}

\maketitle

\section{Introduction}

The micro-gravity environment provides a good platform for studying physics under ideal conditions. Recently, many experiments on cold atoms based on micro-gravity environment have made remarkable progress, such as the cold atom experiments in the micro-gravity environment of the space station \cite{space-01-liu2018orbit, space-02-lachmann2021ultracold, space-03-carollo2022observation, space-04-gaaloul2022space}, and the cold atom interferometer experiment in free fall \cite{space-05-muntinga2013interferometry}. These efforts have provided scientists with a deeper understanding of the Bose-Einstein condensate (BEC), which was theoretical predicted by Bose \cite{bose1924plancks} and Einstein \cite{einstein1924quantum}, experimentally realised in dilute atomic gases \cite{anderson1995}. The interference between two BECs has been observed to characterise its coherent nature \cite{andrews1997observation}. The Gross-Pitaeviskii (GP) equation \cite{gross1961structure, gross1963hydrodynamics, pitaevskii1961vortex, dalfovo1999theory} which is a type of non-linear Schr\"{o}dinger equations, has been the paradigm to explore the coherent non-linear dynamics of BEC \cite{liu2000nonlinear}, especially its macroscopic matter wave interference phenomena.

Many remarkable advances have been made by scientists in recent years regarding the study of BEC \cite{BEC-01-luo2022tunable, BEC-02-zhu2022vortex, BEC-03-luo2022bessel, BEC-04-yasir2022topological, BEC-05-zhu2022manipulating, BEC-06-zhu2022vortex, BEC-07-kengne2021spatiotemporal, BEC-08-zhu2021spin, BEC-09-kengne2021phase, BEC-10-jia2021dissipative, BEC-11-zou2021formation, BEC-12-bai2021nonlinear, BEC-13-wang2021adjustable, BEC-14-li2020research, BEC-15-luo2019three, BEC-16-kengne2018management, BEC-17-yao2018localized, BEC-18-li2018dynamical, BEC-19-li2018weyl, BEC-20-yasir2017spin, BEC-rmp-01-morsch2006dynamics, curved-01-viermann2022quantum, curved-02-tolosa2022curved, curved-03-bilic2013frw, curved-04-jain2007analog, curved-05-barcelo2001analogue, G-01-garraway1995wave, G-02-vangeleyn2014inductive, G-03-dalton2014new, G-04-dalton2017quantum, G-05-sinuco2019microwave, Juze-01-alonso2022cold, Juze-02-kim2021weakly, Juze-03-yin2020localization, Juze-04-liu2017spin, Juze-05-zheng2015topological, Juze-06-han2015supersolid, Guo-01-su2016controllable, Guo-03-liu2013vortex, Guo-04-su2013kibble, Ooi-01-ng2022measuring, Ooi-02-ooi2019molecular, Ooi-03-enaki2015higher, M-01-sakaguchi2022one, M-02-subramaniyan2022interplay, M-03-santos2022effective, M-04-bhat2021modulational, M-05-kengne2021spatiotemporal, M-06-dos2020double, M-07-huang2020shortcuts}. However, the effect of Earth's gravity on BEC is often overlooked. In fact, the effect of Earth's gravity on the evolution of the wave function is not negligible in the usual laboratory environment on the Earth's surface. The temperature of absolute zero is unreachable, and even microscopic particles with temperatures very close to absolute zero (\textit{e.g.} $1nK$) are still in motion. Due to the disordered motion of microscopic particles and the interconversion of kinetic and potential energy, when they are located at different positions of gravitational potential energy, their kinetic energy will also change, which will change their motion, and therefore the particle density distribution will be changed. Meanwhile, since the phase of wave function is related to momentum, and the change of kinetic energy will also cause the change of momentum, so the phase distribution of wave function will also be changed. Therefore, the effect of the Earth's gravity on conventional BEC experiments on the Earth's surface should be considered unless the system is in a micro-gravity environment.

In this paper, the interference of BECs in gravity was investigated by numerically solving the GP equation. Through considering the gravity from the Earth as the external potential, the phenomenon of gravitational transport of microscopic particles of the macroscopic wave function of the condensates was investigated by comparing the related dynamical density distributions, which disappears in the micro-gravity environment that the Earth's gravity is negligible. The effects of other parameters (interaction strength, initial momentum, phase difference \textit{etc.}) were investigated by adjusting the related values. The results here were compared with the experimental results and showed a general agreement. We also found that the macroscopic distribution of the wave function may bring the macroscopic effective momentum and energy, which cause the interference and dynamical evolution of the condensates with zero initial momentum and energy. In addition, the density of interference fringes was found to be related to the interaction strength and initial momentum, which is consistent with the experiments and our physical intuition.

\section{Method: mean-field theory of Bose-Einstein condensate}

Based on the mean-field theory of weakly interacting bosons (dilute Bose gas) developed by Bogoliubov \cite{bogoliubov1947theory, dalfovo1999theory}, the dynamics of Bose-Einstein condensate (BEC) can be described by time-dependent Gross-Pitaevskii (GP) equation \cite{gross1961structure, gross1963hydrodynamics, pitaevskii1961vortex, dalfovo1999theory} that:
\begin{equation} 
\label{eq-gpe-original}
	i \hbar \frac{\partial}{\partial t} \psi = -\frac{\hbar^2}{2 M} \nabla^2 \psi + \frac{4 \pi \hbar^2 a_s}{M} |\psi|^2 \psi + V_{ext} \psi,
\end{equation} 
where $\hbar$ is the reduced Planck's constant, $M$ is the atomic mass, $a_s$ is the $s$-wave scattering length, $V_{ext}$ is the external potential, $\psi$ is the wave function (sometimes named as \textit{order parameter}) which characterises the possibility distribution of particles. 

For simplification, the dimensionless form of GP equation is preferred. Set the time, space, wave function variables $t$, $x$, $\psi$ as:
\begin{equation}
\label{eq-substitution}
	\begin{split}
		t = \tau \alpha,\;\;\; x = \xi \beta,\;\;\; \psi = \phi \gamma,
	\end{split}
\end{equation}
where the $\alpha$, $\beta$, $\gamma$ are the units, while the $\tau$, $\xi$, $\phi$ are the dimensionless variables of time, space, wave function. Then the $\xi$'s Laplace operator can be defined as: $\nabla_{\xi}^2 = \frac{\partial^2}{\partial \xi_1^2} +\frac{\partial^2}{\partial \xi_2^2} +\frac{\partial^2}{\partial \xi_3^2} = \beta^2(\frac{\partial^2}{\partial x_1^2} +\frac{\partial^2}{\partial x_2^2} +\frac{\partial^2}{\partial x_3^2}) = \beta^2 \nabla^2  $. Notice that the original wave function satisfies the normalisation condition: $ N = \int |\psi|^2 dx_1 dx_2 dx_3 = \beta^3 \gamma^2 \int |\phi|^2 d\xi_1 d\xi_2 d\xi_3 $, where $N$ is the number of particles, this integral must be three-dimensional from the dimensional analysis of GP equation. Let the dimensionless wave function $\phi$ be normalised as: $1 = \int |\phi|^2 d\xi_1 d\xi_2 d\xi_3 $, then $\gamma$ can be presented by $\beta$ that: $\gamma = \sqrt{N /  \beta^3} $. So the GP equation becomes:
\begin{equation}
\label{eq-gpe-dimensionless}
	i \frac{\partial}{\partial \tau} \phi = -\frac{\hbar \alpha}{2 M \beta^2} \nabla_{\xi}^2 \phi + \frac{4 \pi N \hbar a_s  \alpha}{M \beta^3} |\phi|^2 \phi + \frac{V_{ext} \alpha}{\hbar} \phi,
\end{equation} 
which is dimensionless and convenient for calculation. If the gravity is non-negligible, the $V_{ext}$ should be non-zero. 

To solve the GP equation, the initial condition should be determined. As a basic hypothesis of quantum mechanics, the single particle's dimensionless wave function should be: $\phi_{single} = e^{\frac{i}{\hbar}S}$, where $S = \int L dt$ is the action as the temporal integral of Lagrangian $L$ \cite{landau2013quantum}. Through the Legendre transform: $L = p v - H $, notice that: $v dt = dx $, assume the momentum $p$ is independent to $x$ while the Hamiltonian $H$ is independent to $t$, this wave function can be rewritten as: $\phi_{single} = e^{\frac{i}{\hbar} \int p dx} e^{-\frac{i}{\hbar} \int H dt} = e^{\frac{i}{\hbar} p x} e^{-\frac{i}{\hbar} H t}  $, where $e^{\frac{i}{\hbar} p x}$ is the usual form of wave function which becomes $e^{\frac{i}{\hbar} \vec{p} \cdot \vec{x}}$ in higher dimensional space, and the $e^{-\frac{i}{\hbar} H t}$ is the so-called time-evolutional term. Let the many-body wave function of the condensate be the product of the distributional function $\phi_{dis}$ and the single particle's wave function $\phi_{single}$ that: $\phi = \phi_{dis} \times \phi_{single}$. Assume that the initial state satisfies the isotropic Gaussian distribution in two-dimensional space, and the particles own the initial momentum $\vec{p}_{ini}$ pointing outward along the radius from the centre, then the initial wave function $\phi_{ini}$ can be written as:
\begin{equation}
\label{eq-initial}
	\begin{split}
		\phi_{ini} &= \phi_{dis} \times \phi_{single}\\
		&= \sqrt{\left(\frac{1}{\sqrt{2 \pi \sigma^2}} \right)^2 e^{-\frac{\xi_1^2 + \xi_2^2}{2 \sigma^2} } } \times e^{\frac{i \beta}{\hbar} p_{ini} \sqrt{\xi_1^2 + \xi_2^2}},
	\end{split}
\end{equation}
where $|\phi_{ini}|^2$ satisfies the normalisation condition $1 = \int |\phi_{ini}|^2 d\xi_1 d\xi_2 $, the $\sigma$ is the inflection point of the two-dimensional isotropic Gaussian distribution $|\phi_{dis}|^2 =  \left(\frac{1}{\sqrt{2 \pi \sigma^2}} \right)^2 e^{-\frac{\xi_1^2 + \xi_2^2}{2 \sigma^2} }$, the $p_{ini} = \sqrt{|\vec{p}_{ini}|^2}$ is the norm of the initial momentum of a single particle of the condensate, the spacial coordinates have been rewritten as $x = \xi \beta$, the time-evolutional term $e^{-\frac{i}{\hbar} H t}$ has been eliminated as $t=0$ for the initial state.

Let the kinetic term's coefficient of the dimensionless GP equation (\ref{eq-gpe-dimensionless}) equals to $1$ that: $\frac{\hbar \alpha}{2 M \beta^2} = 1 $, then temporal unit $\alpha$ can be represented by spacial unit $\beta$ that: $\alpha = \frac{2 M \beta^2}{\hbar}$. And the interaction term's coefficient becomes: $ \frac{4 \pi N \hbar a_s  \alpha}{M \beta^3} = \frac{8 \pi N a_s }{\beta} $, denote it as: $\eta = \frac{8 \pi N a_s }{\beta}$. Here, the free diffusion case is considered, so the external potential equals to $0$ that: $V_{ext}=0$. Then the GP equation becomes:
\begin{equation}
\label{eq-gpe-free-simple}
	i \frac{\partial}{\partial \tau} \phi = - \nabla_{\xi}^2 \phi + \eta |\phi|^2 \phi,
\end{equation} 
where the $\eta$ characterises the interaction strength, and it can be adjusted by adjusting the $s$-wave scattering length $a_s$ by the effects of Feshbach resonance \cite{FR-01-chin2010feshbach, FR-02-feshbach1958unified, FR-03-feshbach1962unified, FR-04-courteille1998observation} in experiments. 

Consider two trapped Bose-Einstein condensates separated with a distance $2 \delta_d$, assume their possibility distribution satisfies the Gaussian distribution in Eq. (\ref{eq-initial}), the initial wave function can be written as:
\begin{equation}
\label{eq-initial-separated}
	\begin{split}
		\phi_0 &= \phi_{ini}(\xi_1 + \delta_d,\; \xi_2) + \phi_{ini}(\xi_1 - \delta_d,\; \xi_2) \\
		&= \sqrt{\left(\frac{1}{\sqrt{2 \pi \sigma^2}} \right)^2 e^{-\frac{(\xi_1 + \delta_d)^2 + \xi_2^2}{2 \sigma^2} } } \times e^{\frac{i \beta}{\hbar} p_{ini} \sqrt{(\xi_1 + \delta_d)^2 + \xi_2^2}} \\
		&+ \sqrt{\left(\frac{1}{\sqrt{2 \pi \sigma^2}} \right)^2 e^{-\frac{(\xi_1 - \delta_d)^2 + \xi_2^2}{2 \sigma^2} } } \times e^{\frac{i \beta}{\hbar} p_{ini} \sqrt{(\xi_1 - \delta_d)^2 + \xi_2^2}},
	\end{split}
\end{equation}
where the condensates are separated on $\xi_1$-axis.
In micro-gravity environment that the gravity from the Earth is counteracted by the space station's orbital motion \cite{space-01-liu2018orbit, space-02-lachmann2021ultracold, space-03-carollo2022observation, space-04-gaaloul2022space} or by the experimental device's free fall \cite{space-05-muntinga2013interferometry}, cancel the external trapping potential and let the condensates diffuse freely, the dynamics of the wave function can be obtained by solving the GP equation (\ref{eq-gpe-free-simple}) with the initial condition in Eq. (\ref{eq-initial-separated}).   

\section{Interference influenced by interaction strength in micro-gravity}

The experiments \cite{space-01-liu2018orbit, space-02-lachmann2021ultracold, space-03-carollo2022observation, space-04-gaaloul2022space, space-05-muntinga2013interferometry} with micro-gravity environment have made great progress for human to explore the pure nature. Micro-gravity environment, usually refers to an environment where the gravity from the Earth is so weak that it can be almost ignored. It can be constructed by the space station moving around the Earth at suitable speed in earth orbit \cite{space-01-liu2018orbit, space-02-lachmann2021ultracold, space-03-carollo2022observation, space-04-gaaloul2022space}, or by the experimental setup in free fall \cite{space-05-muntinga2013interferometry}. All these micro-gravity environments are based on Einstein's equivalence principle \cite{ep-01-einstein1911,ep-02-nordtvedt1968equivalence, ep-03-nordtvedt1968equivalence}. 

To investigate the effects of the interaction strength on the interference pattern of two BECs in micro-gravity (gravity is weak enough to be negligible), consider the dimensionless GP equation (\ref{eq-gpe-free-simple}) with the initial state in Eq. (\ref{eq-initial-separated}). Set the parameters of the initial wave function as: $\sigma = 1$, $\frac{\beta p_{ini}}{\hbar} =1 $, $\delta_d = 2$, modify the GP equation's interaction strength $\eta$, then solve the equation numerically, and the results can be shown as follows.  

\begin{figure}[ ]
	\centering
	\includegraphics[width = 0.45 \textwidth ]{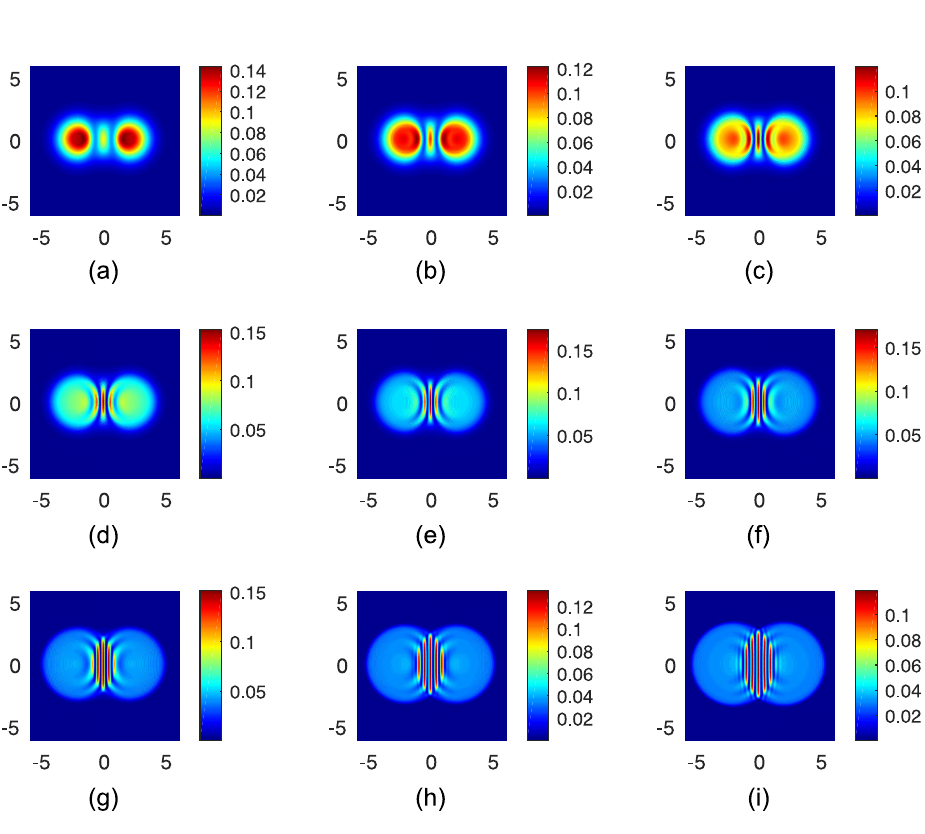}
	\caption{The interference of two Bose-Einstein condensates in two-dimensional space. Here, the parameters are chosen as: $\sigma = 1$, $\frac{\beta p_{ini}}{\hbar} =1 $, $\delta_d = 2$, $\eta=2000$. The evolution of $|\phi|^2$ has been shown in sub-figures, from $\tau = 0.01$ in (a), to $\tau = 0.09$ in (i), the temporal intervals between the neighbouring sub-figures are $0.01$. }
	\label{fig-eta2000}
\end{figure}

From the temporal evolution of the wave function's density distribution $|\phi|^2$ in Fig. \ref{fig-eta2000}, as time increases, the condensate gradually spreads to the surroundings, and interference occurs when the wave functions of two condensates overlap. To find the effect of interaction strength on the interference pattern, modify the value of interaction strength and numerically solve the GP equation, the results have been shown as follows.

\begin{figure}[ ]
	\centering
	\includegraphics[width = 0.45 \textwidth]{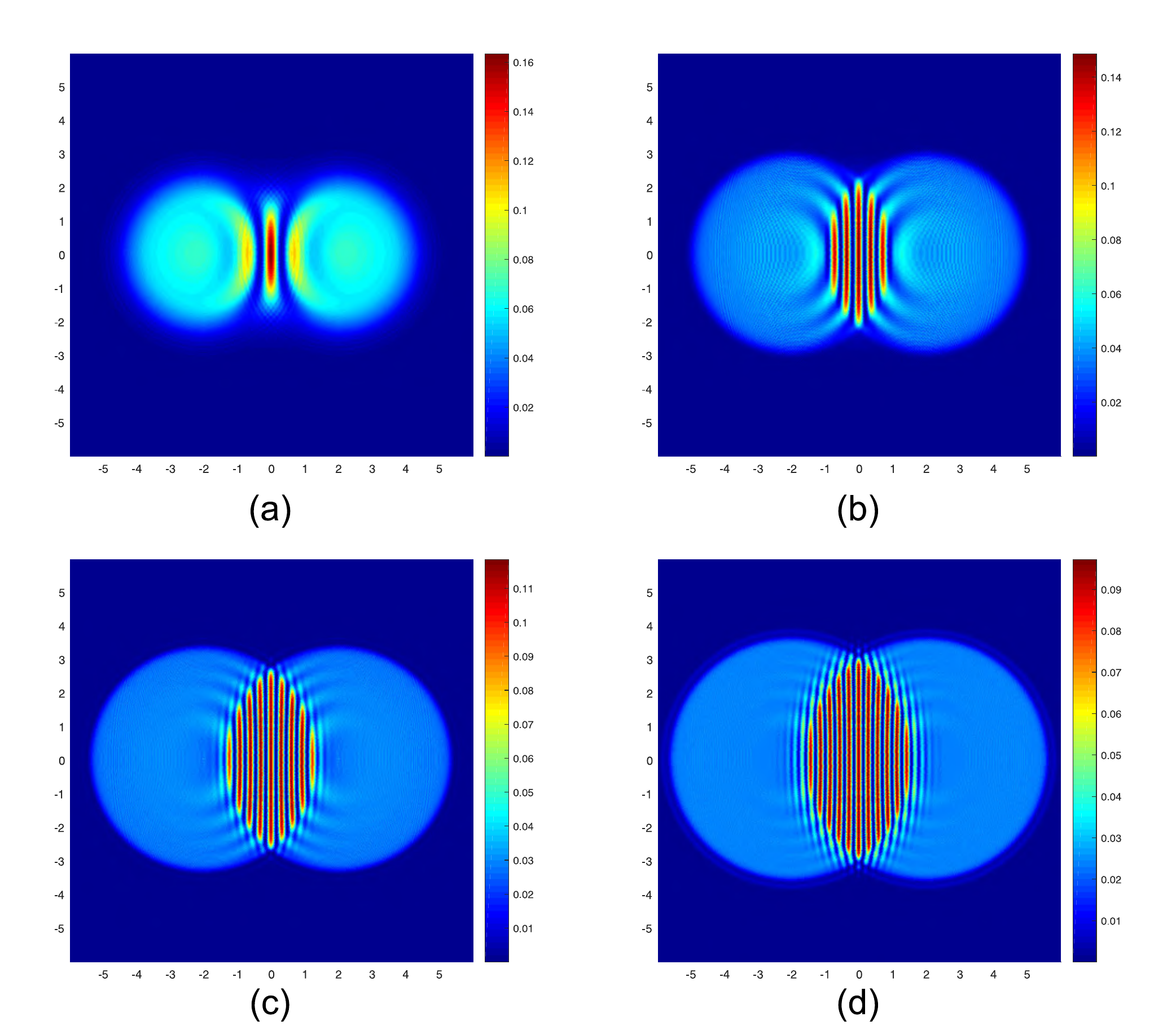}
	\caption{Comparison of the interference fringes with different interaction strengths. Here, the temporal parameter is set as: $\tau=0.06$, the other fixed parameters are set as: $\sigma = 1$, $\frac{\beta p_{ini}}{\hbar} =1 $, $\delta_d = 2$. The interaction strengths are: $\eta = 1000$ in (a), $\eta = 3000$ in (b), $\eta = 5000$ in (c), $\eta = 7000$ in (d). }
	\label{fig-compare-eta}
\end{figure}

Here, the temporal parameter $\tau$ is fixed at $\tau = 0.06$, and the interaction strength $\eta$ is set as $\eta=1000$, $\eta=3000$, $\eta=5000$, $\eta=7000$ in the sub-figures of Fig. \ref{fig-compare-eta} to investigate the effects of varying interaction strength. It shows that, the increased interaction strength will accelerate the speed of the evolution of the condensate. The GP equation describes the weakly repelling many-body system of bosons \cite{gross1961structure}, the larger repelling interaction may cause the faster diffusing of the atomic cloud. At the same time, the interaction energy will convert into kinetic energy $E_k$, which related to the momentum $p = \sqrt{2 M E_k} $, that may decrease the de Broglie wave length $\lambda = h/p $ (here $h$ is the Planck's constant), which  characterises the fringe spacing \cite{andrews1997observation}, so the interference fringes may become denser with increased interaction strength $\eta$ as the Fig. \ref{fig-compare-eta} presents.

\section{Interference influenced by initial momentum in micro-gravity}

\begin{figure}[ ]
	\centering
	\includegraphics[width = 0.45 \textwidth]{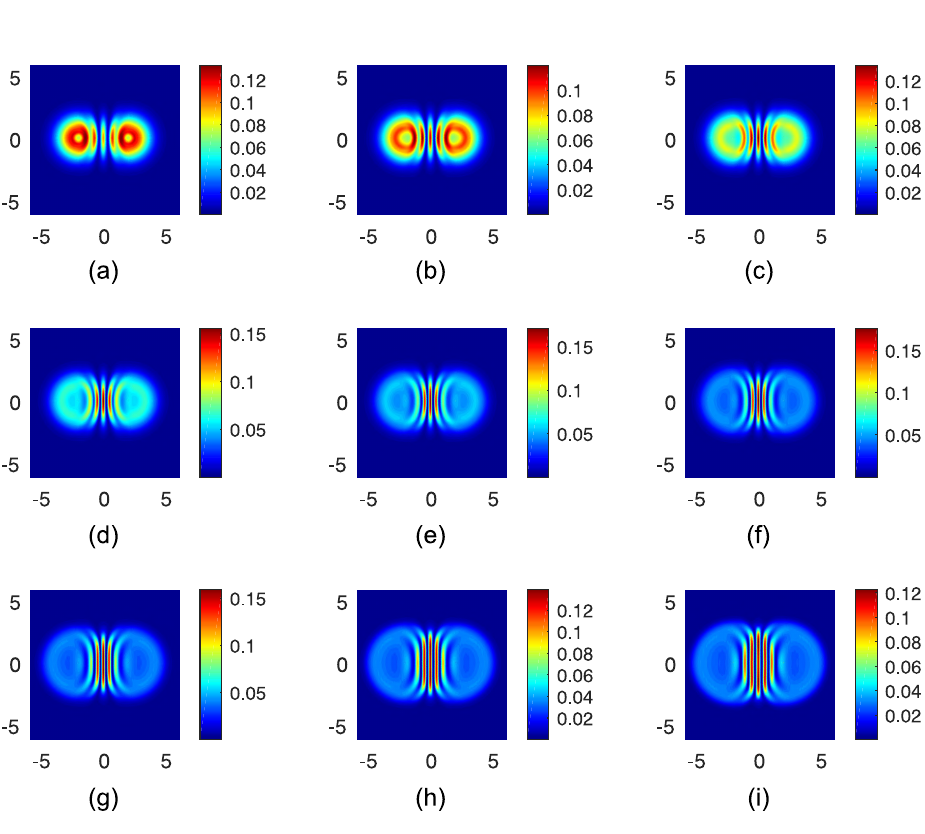}
	\caption{The interference of two Bose-Einstein condensates in two-dimensional space. Here, the parameters are chosen as: $\eta=1000$, $\sigma = 1$, $\delta_d = 2$, $\frac{\beta p_{ini}}{\hbar} = 4 $. The evolution of $|\phi|^2$ has been shown in sub-figures, from $\tau = 0.01$ in (a), to $\tau = 0.09$ in (i), the temporal intervals between the neighbouring sub-figures are $0.01$. }
	\label{fig-k=4}
\end{figure}

\begin{figure}[ ]
	\centering
	\includegraphics[width = 0.45 \textwidth]{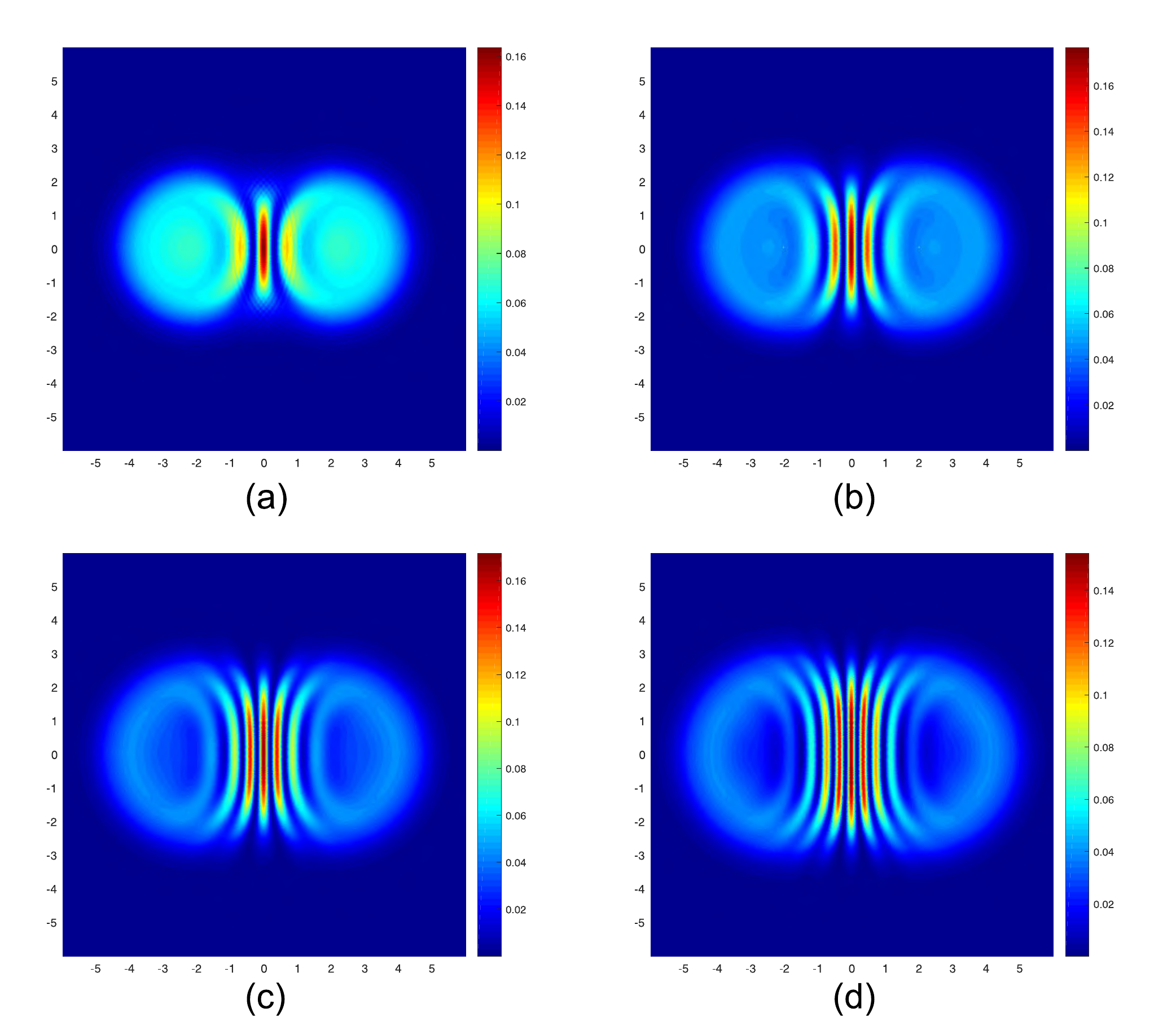}
	\caption{Comparison of the interference fringes with different initial momentum parameter. Here, the temporal parameter is set as: $\tau=0.06$, the other fixed parameters are set as: $\eta=1000$, $\sigma = 1$, $\delta_d = 2$. The initial momentum parameters are: $\frac{\beta p_{ini}}{\hbar} = 1$ in (a), $\frac{\beta p_{ini}}{\hbar} = 3$ in (b), $\frac{\beta p_{ini}}{\hbar} = 5$ in (c), $\frac{\beta p_{ini}}{\hbar} = 7$ in (d). }
	\label{fig-compare-momentum}
\end{figure}

Consider the dimensionless GP equation (\ref{eq-gpe-free-simple}) with the initial wave in Eq. (\ref{eq-initial-separated}), set the parameters as: $\eta = 1000$, $\sigma = 1 $, $\delta_d = 2$, then modify the value of the initial momentum parameter $\frac{\beta p_{ini}}{\hbar} $ which is related to the temperature of the condensate, and numerically solve the GP equation to investigate the the effects of initial momentum on the interference fringes of the condensates in micro-gravity environment that the gravity from the Earth is negligible. The results have been presented in Figs. \ref{fig-k=4} and \ref{fig-compare-momentum}.  

The sub-figures in Fig. \ref{fig-k=4} show the temporal evolution of the density of condensate presented by the wave function's squared norm $|\phi|^2$. To find out the effects of the initial momentum parameter $\frac{\beta p_{ini}}{\hbar}$ on the interference pattern, modify its value and numerically solve the GP equation, the results are presented in Fig. \ref{fig-compare-momentum}. It shows that, the larger initial momentum parameter may cause the denser interference fringes between the condensates. These patterns are different from the patterns caused by large interaction strength $\eta$ in Fig. \ref{fig-compare-eta}. This phase varying caused by initial momentum depends on distance, while the phase varying caused by interaction strength is also related to the condensate's density $|\phi|^2$, their temporal evolutions are different, so their interference patterns are different too. 

\section{Comparison of the influence from interaction and initial momentum in micro-gravity}

The evolution of the interference patterns caused by modified interaction strength or initial momentum parameter are different. Here, these two parameters in Eqs. (\ref{eq-gpe-free-simple}) and (\ref{eq-initial-separated}) will be set as zero respectively to observe the differences between the effects of atomic interaction and initial momentum in micro-gravity environment that the gravity from the Earth is negligible.  

\begin{figure}[ ]
	\centering
	\includegraphics[width = 0.45 \textwidth ]{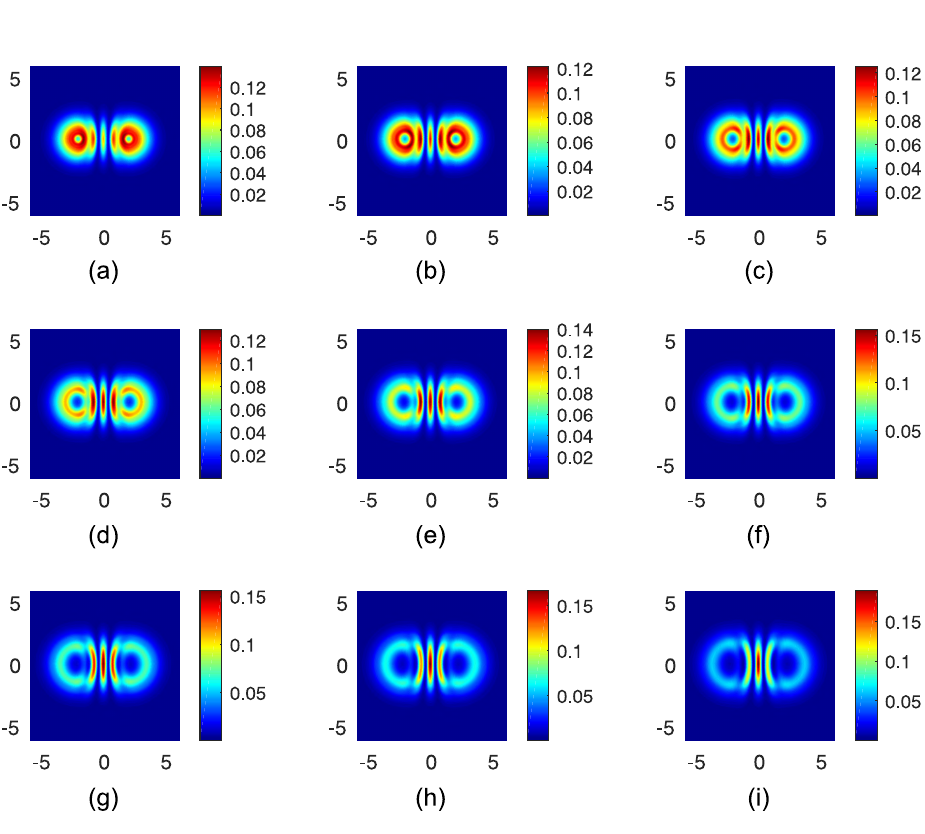}
	\caption{The interference of two Bose-Einstein condensates in two-dimensional space. Here, the parameters are chosen as: $\sigma = 1$, $\delta_d = 2$, $\eta=0$, $\frac{\beta p_{ini}}{\hbar} =4 $. The evolution of $|\phi|^2$ has been shown in sub-figures, from $\tau = 0.01$ in (a), to $\tau = 0.09$ in (i), the temporal intervals between the neighbouring sub-figures are $0.01$. }
	\label{fig-eta0-k4}
\end{figure}

In Fig. \ref{fig-eta0-k4}, the interaction strength is set as: $\eta = 0$, while the initial momentum parameter is set as: $\frac{\beta p_{ini}}{\hbar} = 4$. Here the GP equation is reduced into the Schr\"{o}dinger equation, since the non-linear term $\eta |\phi|^2 \phi$ has disappeared. Without the atomic interaction, the condensate's macroscopic wave function becomes just the superposition of all the single particle's wave function. Consequently, the interference patterns caused by the non-linear effects has also disappeared, while the interference patterns caused by phase difference from the initial momentum remains.

\begin{figure}[ ]
	\centering
	\includegraphics[width = 0.45 \textwidth ]{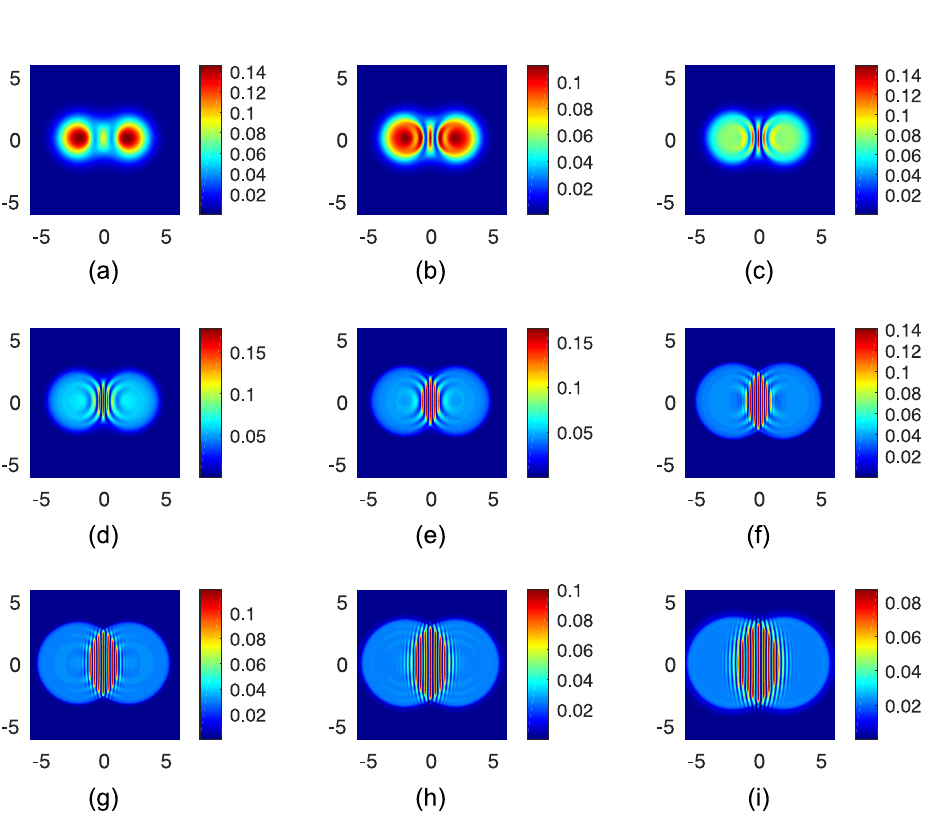}
	\caption{The interference of two Bose-Einstein condensates in two-dimensional space. Here, the parameters are chosen as: $\sigma = 1$, $\delta_d = 2$, $\eta=4000$, $\frac{\beta p_{ini}}{\hbar} =0 $. The evolution of $|\phi|^2$ has been shown in sub-figures, from $\tau = 0.01$ in (a), to $\tau = 0.09$ in (i), the temporal intervals between the neighbouring sub-figures are $0.01$.}
	\label{fig-eta4000-k0}
\end{figure}

The results in Fig. \ref{fig-eta4000-k0} are obtained with the interaction strength: $\eta = 4000$, and the initial momentum parameter: $\frac{\beta p_{ini}}{\hbar} = 0$. Here, the initial momentum of every single particle of the condensate is zero, the diffusion and the interference are caused by the non-linear interaction term $\eta |\phi|^2 \phi$. In this process, the atomic repulsive interaction may push the atoms to move outward, which presented by the diffusion of the condensate. At the same time, the interaction potential energy will convert into the kinetic energy which correspond to the momentum \textit{i.e.} the phase of the wave function, then the non-linear interference pattern appears.

\begin{figure}[ ]
	\centering
	\includegraphics[width = 0.45 \textwidth ]{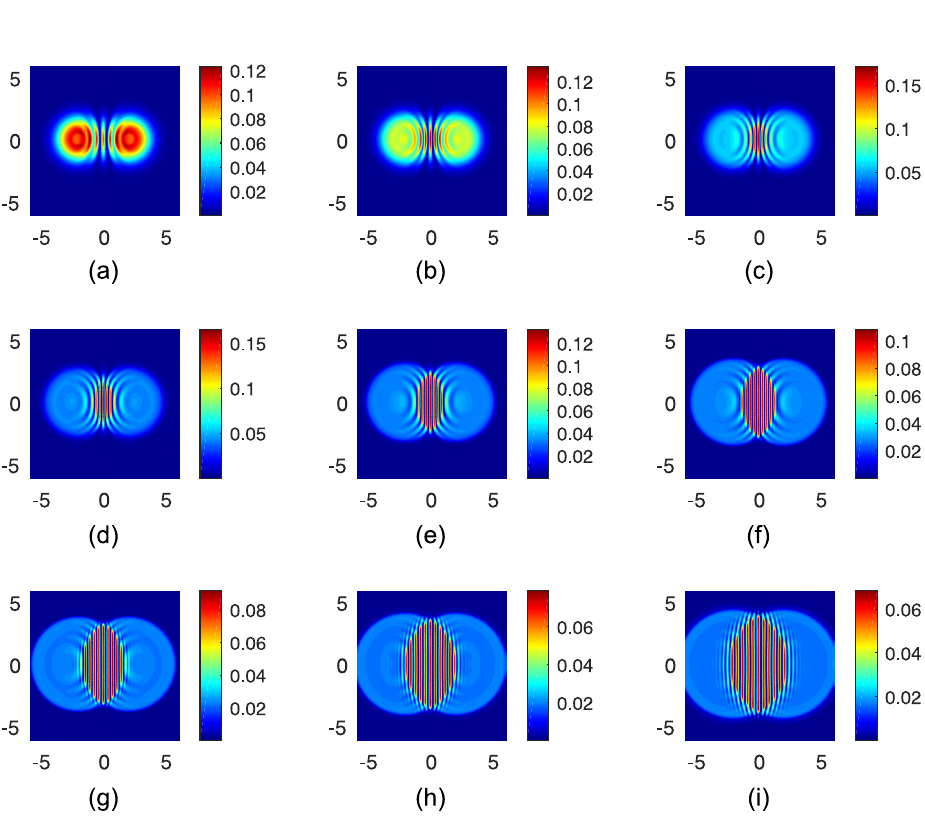}
	\caption{The interference of two Bose-Einstein condensates in two-dimensional space. Here, the parameters are chosen as: $\sigma = 1$, $\delta_d = 2$, $\eta=4000$, $\frac{\beta p_{ini}}{\hbar} =4 $. The evolution of $|\phi|^2$ has been shown in sub-figures, from $\tau = 0.01$ in (a), to $\tau = 0.09$ in (i), the temporal intervals between the neighbouring sub-figures are $0.01$.}
	\label{fig-eta4000-k4}
\end{figure}

In Fig. \ref{fig-eta4000-k4}, the interaction strength is set as: $\eta = 4000$, while the initial momentum parameter is set as: $\frac{\beta p_{ini}}{\hbar} = 4$. Here, the interference patterns are cause by both the atomic interaction and the initial momentum. At the beginning, the effect of the non-linear interaction term is weak, and the initial momentum supports the most of the interference fringes, so these patterns are similar to these in Fig. \ref{fig-eta0-k4}. As the time increases, more and more potential energy has been converted into the kinetic energy which correspond to the momentum, the non-linear interaction term supports the most of the interference fringes, these patterns become similar to these in Fig. \ref{fig-eta4000-k0}, and the interference fringes are denser because the larger initial momentum brings the larger final momentum.

\begin{figure}[ ]
	\centering
	\includegraphics[width = 0.45 \textwidth ]{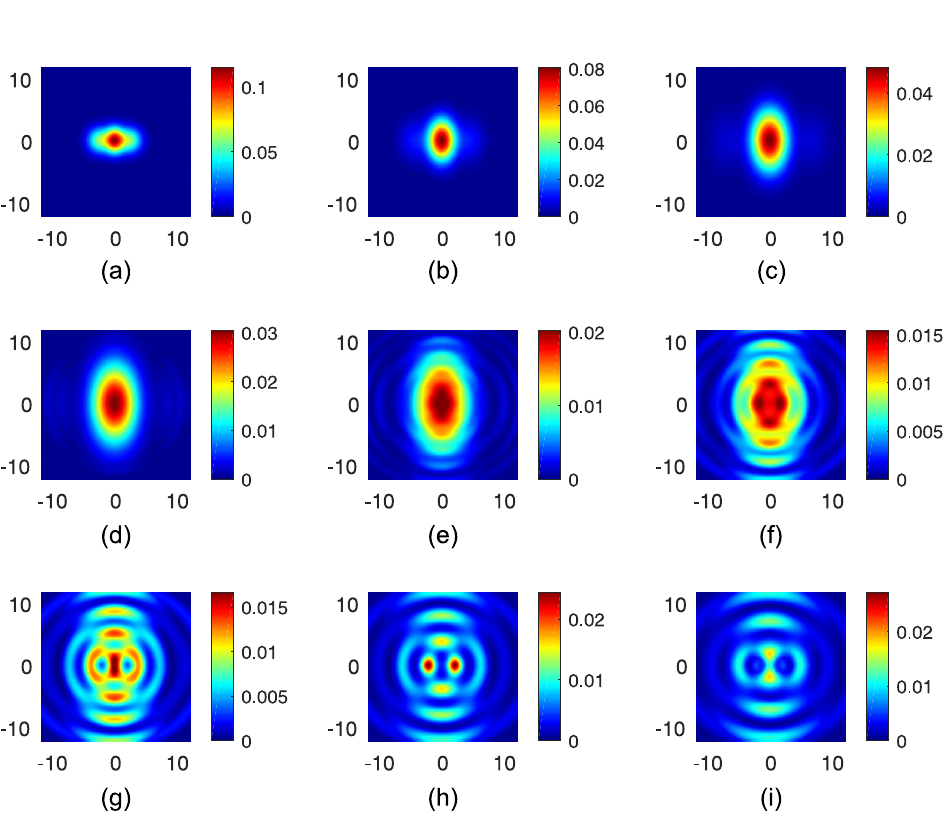}
	\caption{The interference of two Bose-Einstein condensates in two-dimensional space. Here, the parameters are chosen as: $\sigma = 1$, $\delta_d = 2$, $\eta=0$, $\frac{\beta p_{ini}}{\hbar} =0 $. The evolution of $|\phi|^2$ has been shown in sub-figures, from $\tau = 1$ in (a), to $\tau = 9$ in (i), the temporal intervals between the neighbouring sub-figures are $1$.}
	\label{fig-eta0-k0-Lt10}
\end{figure}

When the interaction strength and the initial momentum parameter are both zero, the evolution of the condensate becomes much slower than before, to observe the patterns, the temporal scale should be enlarged. In Fig. \ref{fig-eta0-k0-Lt10}, the interaction and initial momentum parameters are: $\eta=0$, $\frac{\beta p_{ini}}{\hbar} = 0$, the region of time satisfies: $\tau \in [1,9]$, which is $100$ times larger than before. Here, the initial total energy is zero, as the kinetic energy which related to the initial momentum and the potential energy are both zero. So why the condensate's macroscopic density distribution $|\phi|^2$ is not static? The key point is the macroscopic Gaussian distribution. Return to the initial wave function in Eq. (\ref{eq-initial-different momentum}), remember that the wave function should satisfies: $\phi \propto e^{\frac{i}{\hbar} \int \vec{p} \cdot d\vec{x} } = e^{\frac{i \beta}{\hbar} \int \vec{p} \cdot d\vec{\xi} } $, when the initial momentum equals to zero that: $p_{ini} = 0$, the relations above implies that: $ \frac{i \beta}{\hbar} \int \vec{p} \cdot d\vec{\xi} = -\frac{\xi^2}{2 \sigma^2}$, so the momentum must satisfies: $\vec{p} = \frac{i \hbar }{\sigma^2 \beta } \vec{\xi}$. This momentum is a macroscopic effective momentum which emerges from the many-body macroscopic Gauss-type wave function, we designate it as:
\begin{equation}
	\vec{p}_{mac} = \frac{i \hbar }{\sigma^2 \beta } \vec{\xi},
\end{equation}
which is purely imaginary valued and proportional to the spacial coordinate $\vec{\xi}$. Then the related macroscopic effective Hamiltonian can be presented as:
\begin{equation}
	\begin{split}
		H_{mac} = \frac{p_{mac}^2}{2 M_{mac}} = -\frac{\hbar^2}{2 M_{mac} \sigma^4 \beta^2} \xi^2,
	\end{split}
\end{equation}
where $M_{mac}$ is the macroscopic effective mass. This Hamiltonian characterises the non-zero effective energy of the initial state, which implies the dynamical evolution of the condensate's macroscopic density distribution, as Fig. \ref{fig-eta0-k0-Lt10} has shown.

\begin{figure}
	\centering
	\includegraphics[width = 0.45 \textwidth]{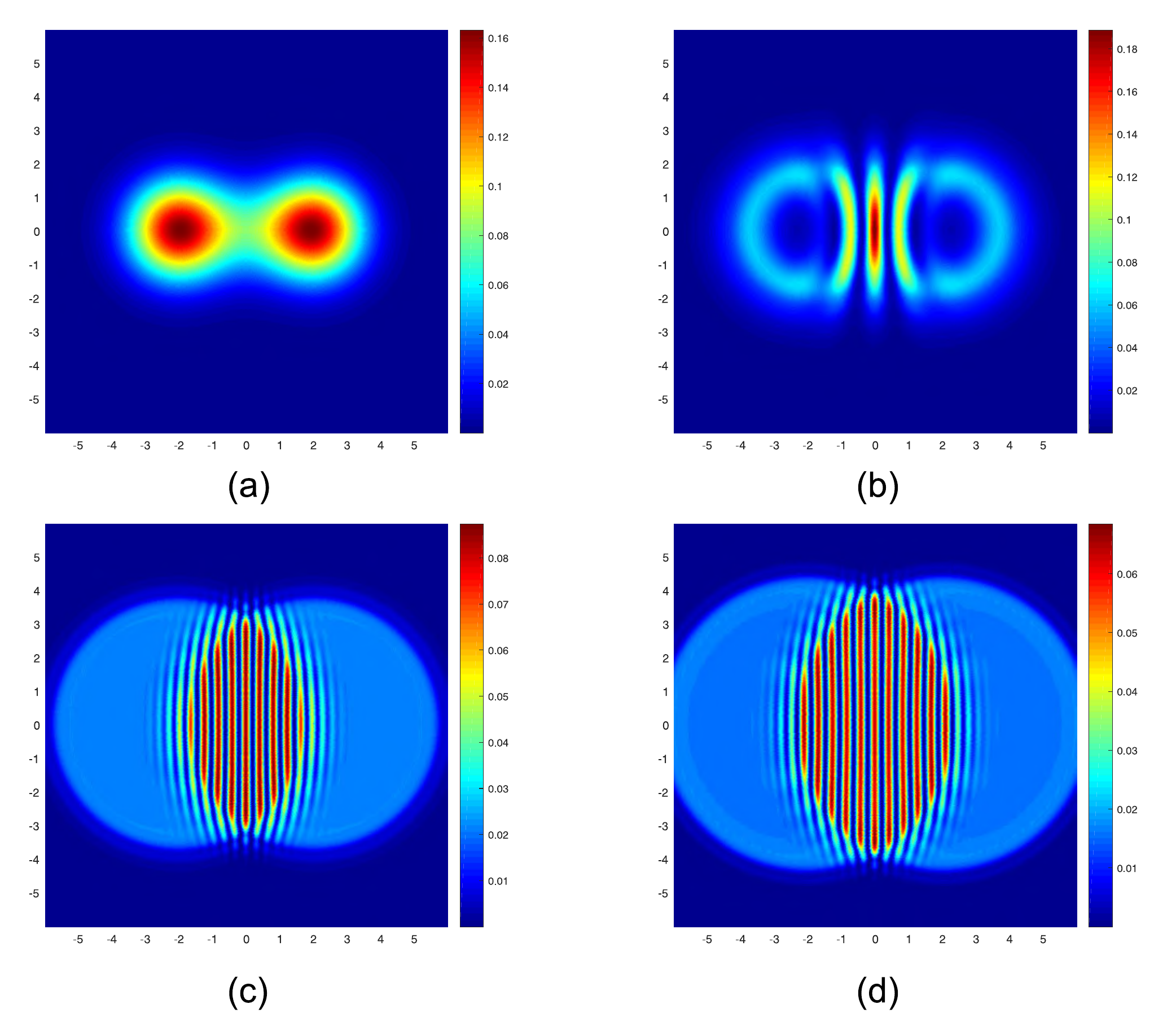}
	\caption{Comparison of the interference fringes with different interaction strength or initial momentum parameter. Here, the temporal parameter is set as: $\tau=0.09$, the other fixed parameters are set as: $\sigma = 1$, $\delta_d = 2$. The interaction strengths and initial momentum parameters are: $\eta=0$, $\frac{\beta p_{ini}}{\hbar}=0$ in (a); $\eta=0$, $\frac{\beta p_{ini}}{\hbar}=4$ in (b); $\eta=4000$, $\frac{\beta p_{ini}}{\hbar}=0$ in (c); $\eta=4000$, $\frac{\beta p_{ini}}{\hbar}=4$ in (d).  }
	\label{fig-compare-interaction-momentum}
\end{figure}

In Fig. \ref{fig-compare-interaction-momentum}, the comparison of the interaction fringes with or without the interaction or initial momentum are shown. It is obviously that the interaction strength may influence the speed of the evolution of the macroscopic wave function, and the density of interference fringes. The larger interaction strength may cause the faster evolution of the wave function and the denser interference fringes. The initial momentum may influence the density of the interference fringes, that the larger initial momentum may cause the denser interference fringes.

\section{Interference influenced by gravitational strength}

If the condensates placed in the usual experimental environment on the Earth's surface instead of the micro-gravity environment, the gravitational effects should be taken into account. 

Consider the GP equation (\ref{eq-gpe-dimensionless}) with gravitational potential as the external potential, when the direction of the gravity is parallel to the $\xi_1$-axis, the gravitational potential becomes: $V_{ext} = M g_e x_1 = M g_e \beta \xi_1 $, where $M$ is the atomic mass, $g_e$ is the gravitational acceleration on the Earth's surface. Then the dimensionless gravitational parameter can be designated as: $\eta_g = \frac{M g_e \alpha \beta}{\hbar} $, set the zero point of the gravitational potential energy as $\xi_1 = 0$, and the GP equation becomes: 
\begin{equation}
\label{eq-gpe-gravity}
	i \frac{\partial}{\partial \tau} \phi = - \nabla_{\xi}^2 \phi + \eta |\phi|^2 \phi + \eta_g \xi_1 \phi,
\end{equation}
which characterise the dynamical evolution of the macroscopic wave function of the Bose-Einstein condensates with gravitational effects. Numerically solve this equation with the initial state in Eq. (\ref{eq-initial-separated}), the evolutions of the macroscopic wave functions can be displayed as follows.  

\begin{figure}[ ]
	\centering
	\includegraphics[width = 0.45 \textwidth ]{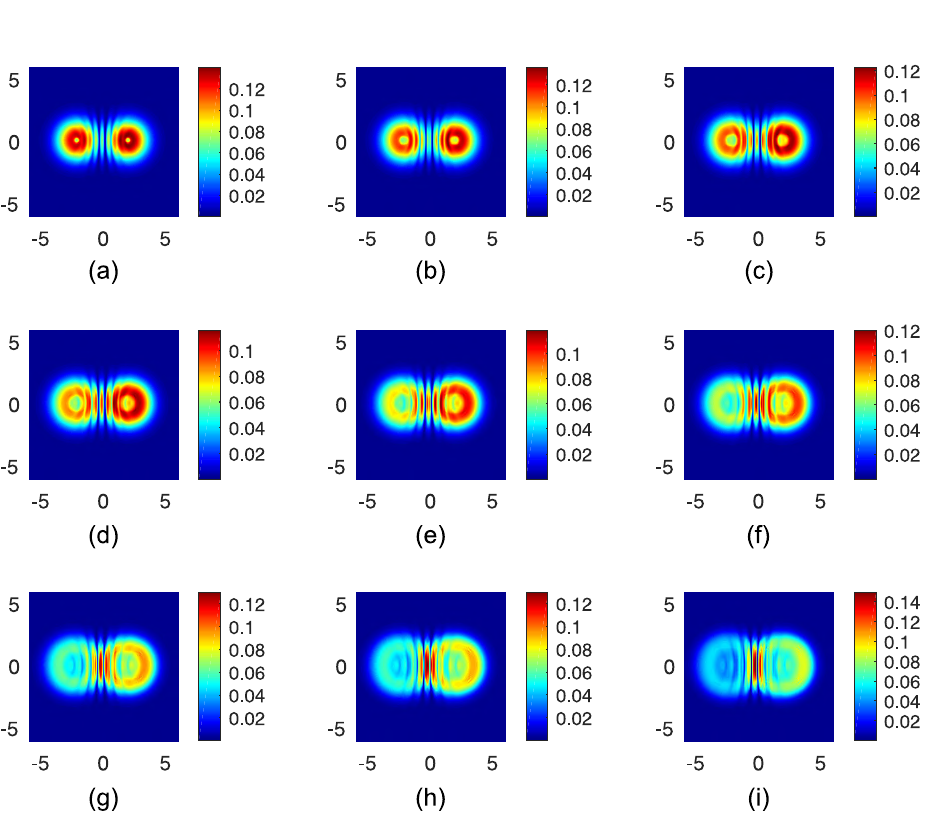}
	\caption{The interference of two Bose-Einstein condensates in two-dimensional space with the gravitational potential along $\xi_1$-axis. Here, the parameters are chosen as: $\sigma = 1$, $\delta_d = 2$, $\eta=3000$, $\frac{\beta p_{ini}}{\hbar} =7 $, $\eta_g=160$. The evolution of $|\phi|^2$ has been shown in sub-figures, from $\tau = 0.004$ in (a), to $\tau = 0.036$ in (i), the temporal intervals between the neighbouring sub-figures are $0.004$. }
	\label{fig-eta_g=160-density}
\end{figure}

\begin{figure}[ ]
	\centering
	\includegraphics[width = 0.45 \textwidth ]{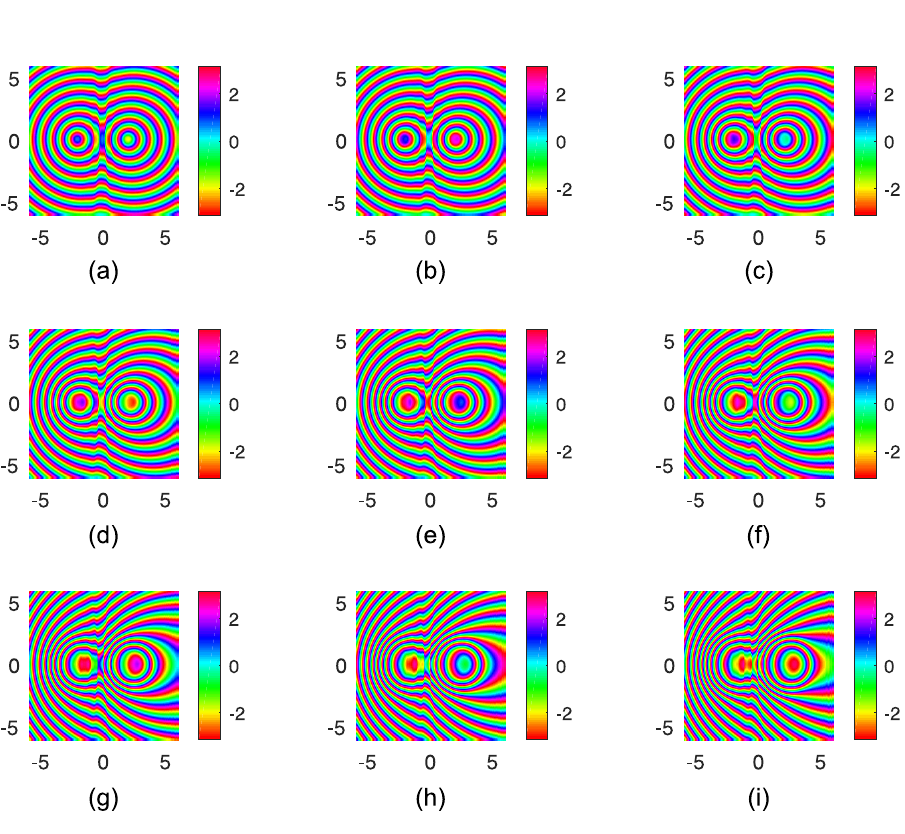}
	\caption{The evolution of the distribution of the angular phase of the complex valued macroscopic wave function $\phi$ of the condensates, which related to the density distribution in Fig. \ref{fig-eta_g=160-density}. The parameters are same with those in Fig. \ref{fig-eta_g=160-density}. }
	\label{fig-eta_g=160-phase}
\end{figure} 

In Fig. \ref{fig-eta_g=160-density}, the temporal evolution of the density $|\phi|^2$ of the macroscopic wave function of the condensates has been shown, where the gravitational parameter is set as: $\eta_g = 160$. Influenced by the gravitational potential, the density becomes greater at where the gravitational potential energy is higher, \textit{i.e.} the particles tend to move towards places with higher gravitational potential energy. At the same time, the gravity also makes the wave function's angular phase varying faster, which is shown in Fig. \ref{fig-eta_g=160-phase}. 

\begin{figure}[ ]
	\centering
	\includegraphics[width = 0.45 \textwidth]{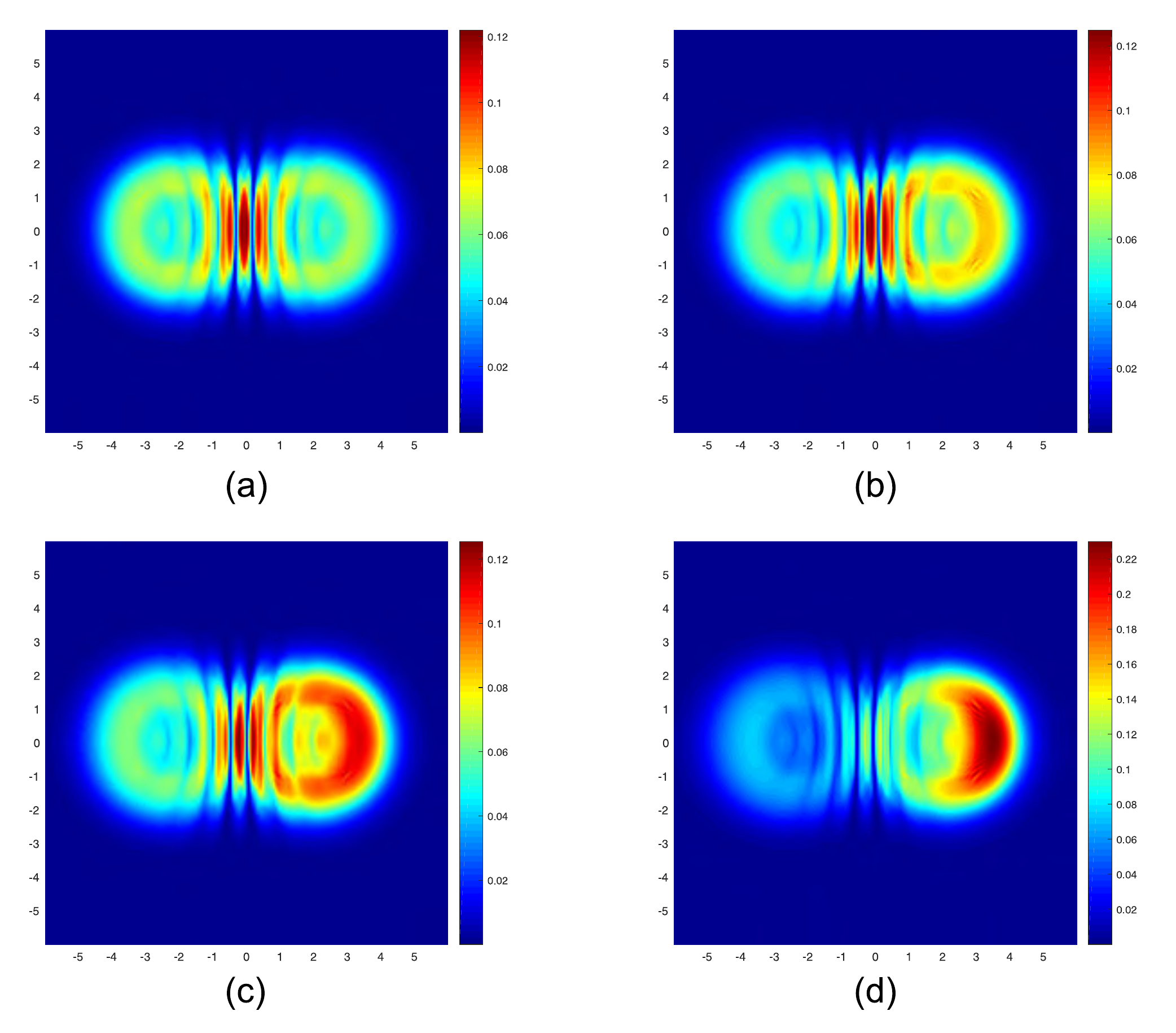}
	\caption{Comparison of the interference fringes with different gravitational potential along $\xi_1$-axis. Here, the temporal parameter is set as: $\tau=0.028$, the other fixed parameters are set as: $\sigma = 1$, $\delta_d = 2$, $\eta = 3000$, $\frac{\beta p_{ini}}{\hbar}=7$. The gravitational parameters are: $\eta_g=0$ in (a), $\eta_g=100$ in (b), $\eta_g=200$ in (c), $\eta_g=300$ in (d). }
	\label{fig-compare-gravity-by density}
\end{figure}

\begin{figure}[ ]
	\centering
	\includegraphics[width = 0.45 \textwidth]{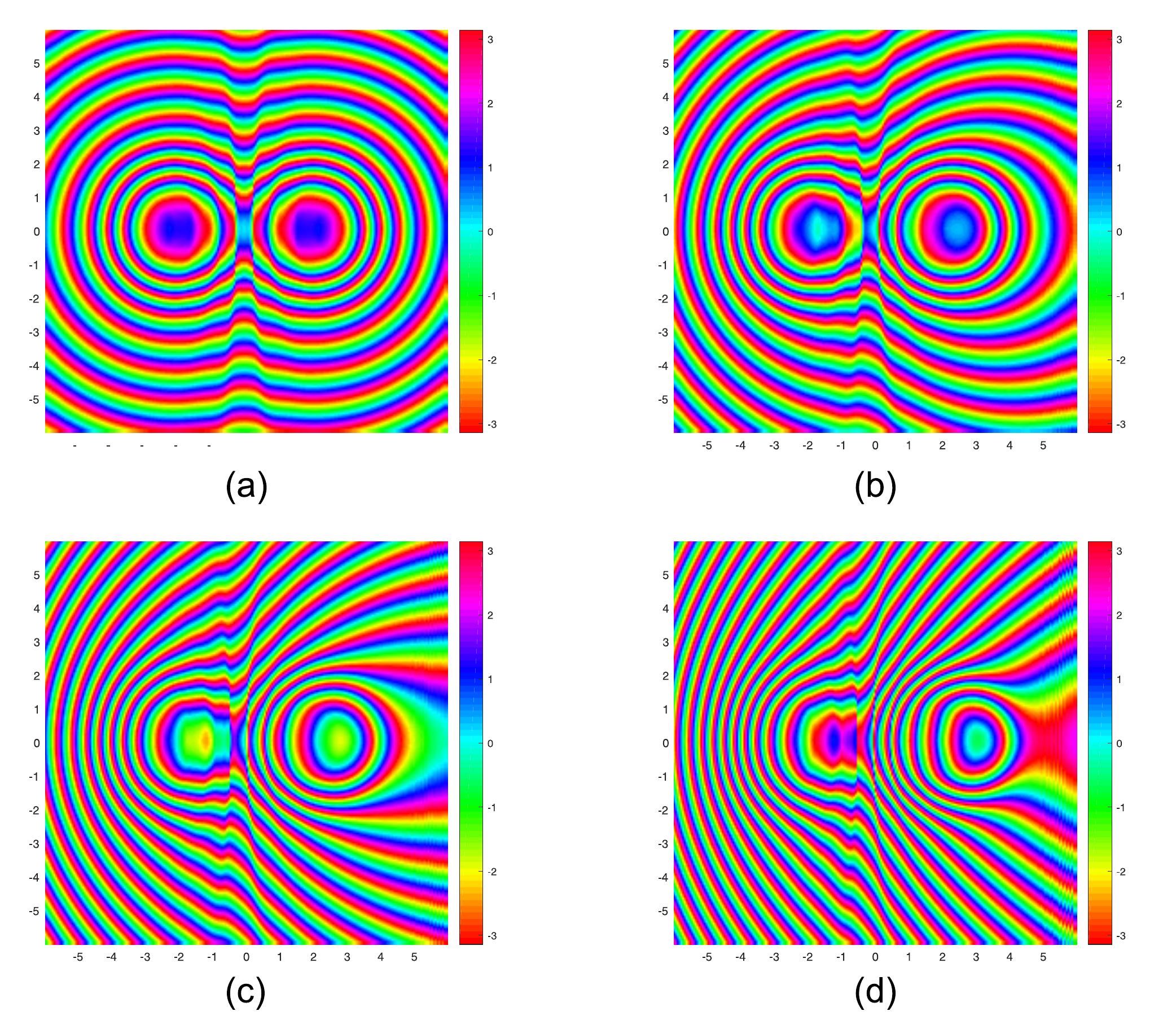}
	\caption{Comparison of the angular phase distribution of the wave function of the interference fringes with different gravitational parameter, which related to the density distribution in Fig. \ref{fig-compare-gravity-by density}. The parameters are same with those in Fig. \ref{fig-compare-gravity-by density}. }
	\label{fig-compare-gravity-by phase}
\end{figure}

To explore the gravitational effects on BECs, the density and phase distributions with different gravitational parameter are shown in Figs. \ref{fig-compare-gravity-by density} and \ref{fig-compare-gravity-by phase} for comparison. The results show that, as the gravitational parameter increases, the related effects become more obvious. The particles may tend to move towards places with higher gravitational potential energy as the wave function's density distribution shows in Fig. \ref{fig-compare-gravity-by density}. The phase may vary faster with the larger gravitational parameter as the wave function's phase distribution shows in Fig. \ref{fig-compare-gravity-by phase}. 

When the direction of the
gravity is parallel to the $\xi_2$-axis, the GP equation becomes:
\begin{equation}
\label{eq-gpe-gravity-y-direction}
	i \frac{\partial}{\partial \tau} \phi = - \nabla_{\xi}^2 \phi + \eta |\phi|^2 \phi + \eta_g \xi_2 \phi,
\end{equation}
where the gravitational parameter is: $\eta_g = \frac{M g_e \alpha \beta}{\hbar} $, the zero point of the gravitational potential energy is set as $\xi_2 = 0$. Numerically solve this equation with the initial state in Eq. (\ref{eq-initial-separated}), the evolution of the macroscopic wave function can be achieved. 

\begin{figure}[ ]
	\centering
	\includegraphics[width = 0.45 \textwidth ]{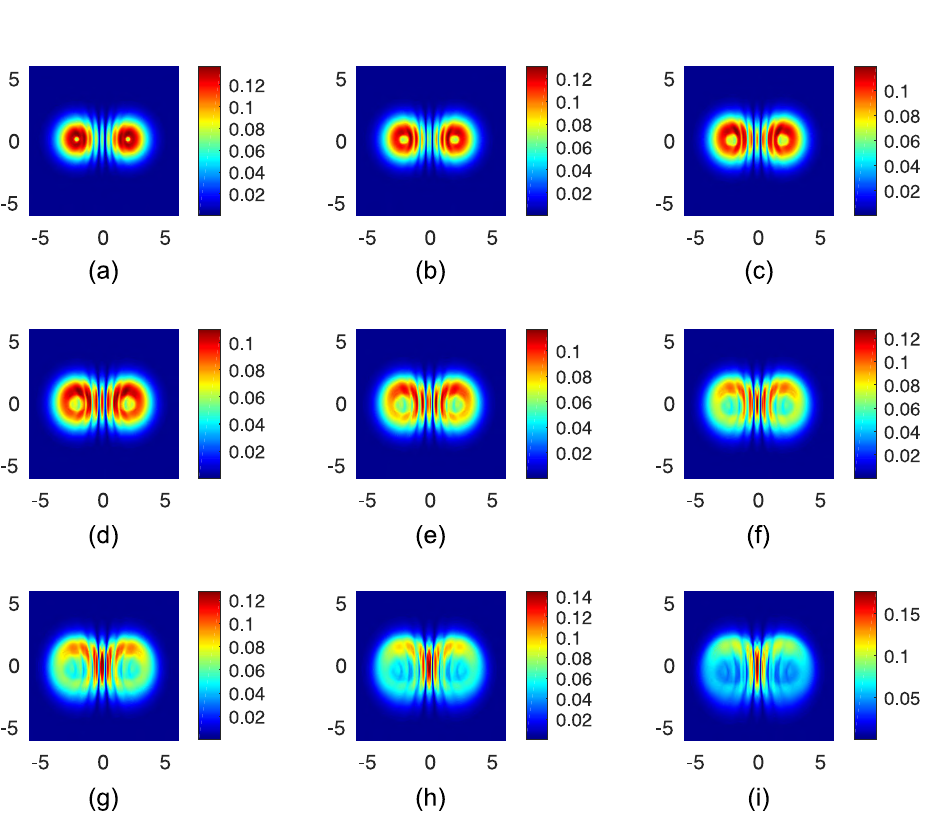}
	\caption{The interference of two Bose-Einstein condensates in two-dimensional space with the gravitational potential along $\xi_2$-axis. Here, the parameters are chosen as: $\sigma = 1$, $\delta_d = 2$, $\eta=3000$, $\frac{\beta p_{ini}}{\hbar} =7 $, $\eta_g=500$. The evolution of $|\phi|^2$ has been shown from $\tau = 0.004$ to $\tau = 0.036$, the temporal intervals between the neighbouring sub-figures are $0.004$. }
	\label{fig-eta_g=500-density}
\end{figure} 

\begin{figure}[ ]
	\centering
	\includegraphics[width = 0.45 \textwidth ]{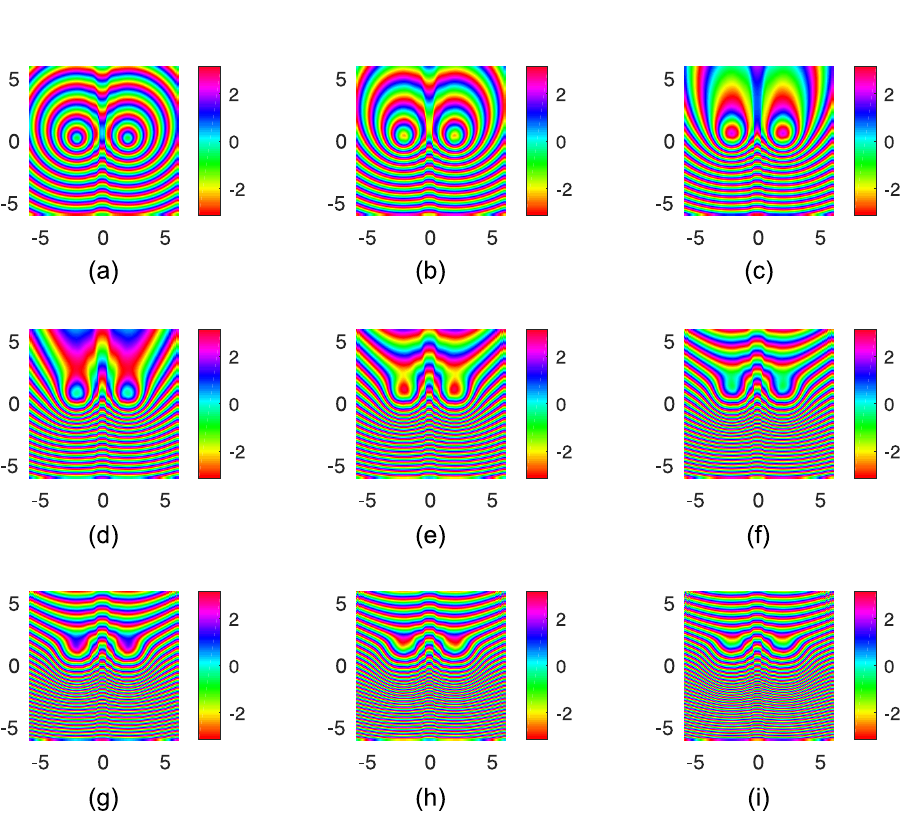}
	\caption{The evolution of the distribution of the angular phase of the complex valued macroscopic wave function $\phi$ of the condensates, which related to the density distribution in Fig. \ref{fig-eta_g=500-density}. The parameters are same with those in Fig. \ref{fig-eta_g=500-density}. }
	\label{fig-eta_g=500-phase}
\end{figure} 

In Figs. \ref{fig-eta_g=500-density} and \ref{fig-eta_g=500-phase}, the temporal evolution of the density and phase distribution of the macroscopic wave function of the BECs with the gravitational potential along $\xi_2$-axis have been presented. It is obvious that the gravitational potential along $\xi_2$-axis may cause the particles moving towards the place with higher gravitational potential energy, which is similar to the case that the gravitational potential is parallel to the $\xi_1$-axis in Fig. \ref{fig-eta_g=160-density}. But the this phenomenon caused by the gravity along $\xi_2$-axis is not very significant compared with the result caused by the gravity along $\xi_1$-axis. This gravity along $\xi_2$-axis also causes the faster varying of the phase of the wave function at the place with the higher gravitational potential, as the gravity along $\xi_1$-axis does in Fig. \ref{fig-eta_g=160-phase}. 

\begin{figure}[ ]
	\centering
	\includegraphics[width = 0.45 \textwidth]{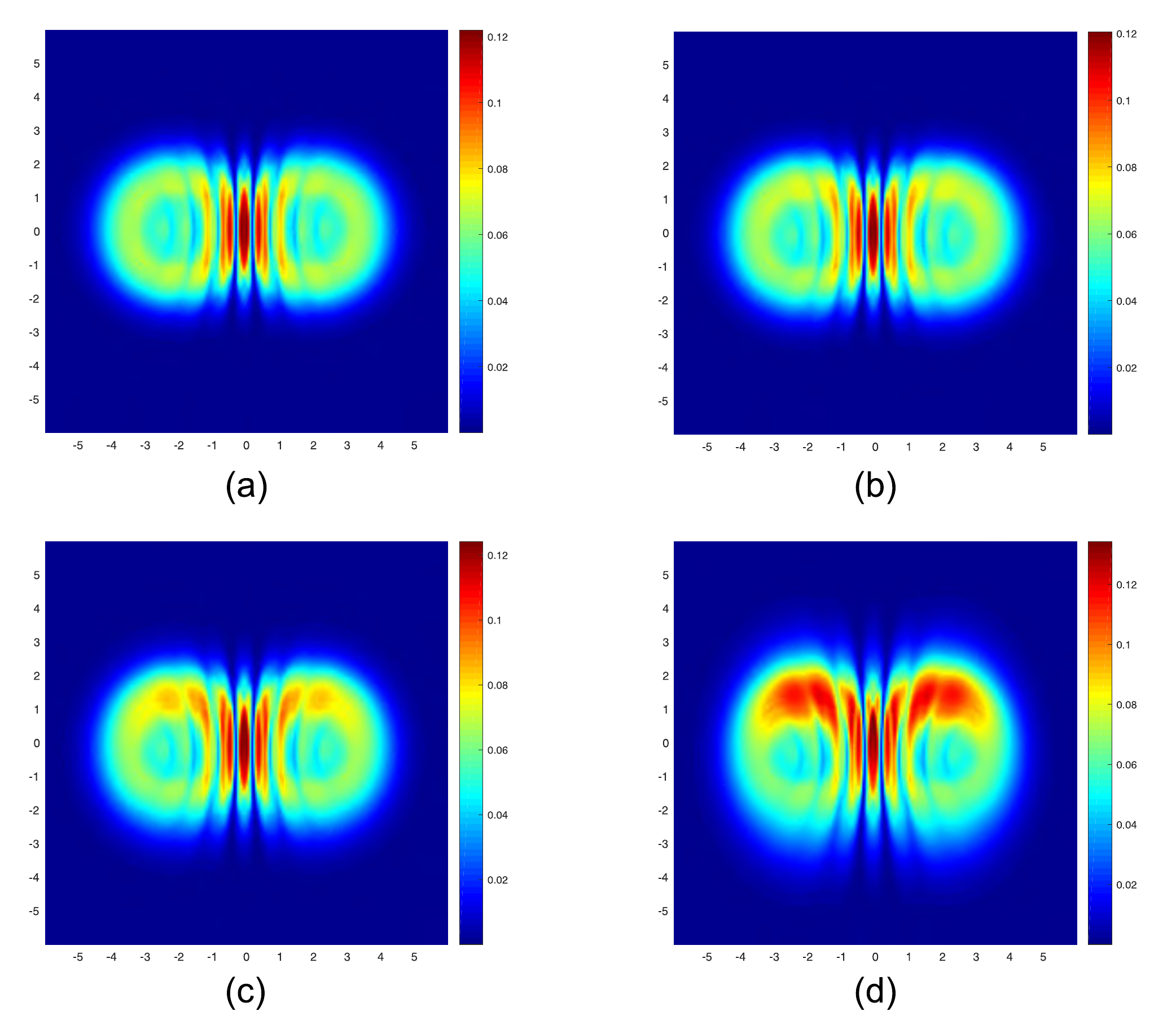}
	\caption{Comparison of the interference fringes with different gravitational potential along $\xi_2$-axis. Here, the temporal parameter is set as: $\tau=0.028$, the other fixed parameters are set as: $\sigma = 1$, $\delta_d = 2$, $\eta = 3000$, $\frac{\beta p_{ini}}{\hbar}=7$. The gravitational parameters are: $\eta_g=0$ in (a), $\eta_g=200$ in (b), $\eta_g=400$ in (c), $\eta_g=600$ in (d). }
	\label{fig-compare-gravity-y-density}
\end{figure}

\begin{figure}[ ]
	\centering
	\includegraphics[width = 0.45 \textwidth]{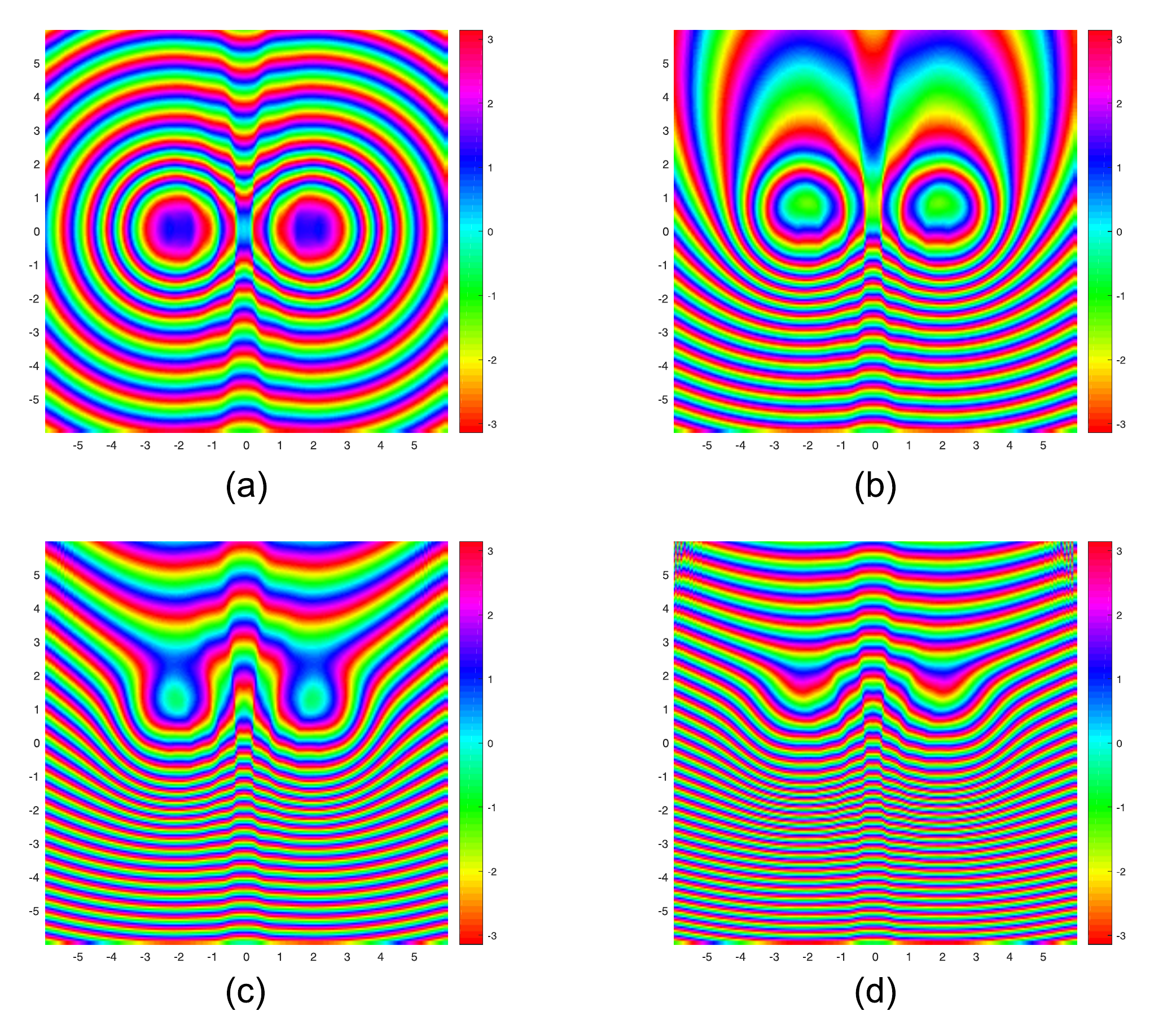}
	\caption{Comparison of the angular phase distribution of the wave function of the interference fringes with different gravitational parameter, which related to the density distribution in Fig. \ref{fig-compare-gravity-y-density}. The parameters are same with those in Fig. \ref{fig-compare-gravity-y-density}. }
	\label{fig-compare-gravity-y-phase}
\end{figure}

To investigate the gravitational effects in other direction, the density and phase distributions of the wave function of the condensates with different gravitational potential along $\xi_2$-axis have been presented in Figs. \ref{fig-compare-gravity-y-density} and \ref{fig-compare-gravity-y-phase}. As the gravitational parameter's value increases, the effects become more obvious. The larger gravitational parameter makes more particles moving to place with higher gravitational potential energy as the density distribution in Fig. \ref{fig-compare-gravity-y-density} shows. At the same time, the phase may vary faster at the place with higher gravitational potential energy as the phase distribution in Fig. \ref{fig-compare-gravity-y-phase} shows. The reasoning is that: the higher gravitational potential energy means higher total energy, \textit{i.e.} the larger Hamiltonian $H$ in $\phi \propto e^{-\frac{i H t}{\hbar}}$ makes the phase of wave function varying faster. 

The gravitational potential can influence both the density and the phase distribution of the macroscopic wave function of BECs. This effect is often neglected in theoretical investigations, but it is not negligible in experiments. Actually, it is hard to construct absolute two-dimensional system experimentally, so the experiments on the Earth's surface must be disturbed by gravity. Because of this, the experiments \cite{space-01-liu2018orbit, space-02-lachmann2021ultracold, space-03-carollo2022observation, space-04-gaaloul2022space, space-05-muntinga2013interferometry} in the microgravity environment of the space station are so precious, which could explore the true nature of nature.

\section{Interference with experimental data}

Consider the experimental data in Ref. \cite{andrews1997observation}, for the sodium-23 ($^{23}\mathrm{Na}$) atoms, the atomic mass: $M = 23 \times 1.66 \times 10^{-27}kg = 3.818 \times 10^{-26} kg$, the $s$-wave scattering length: $a_s = 2.75 \times 10^{-9}m$ \cite{denschlag2000generating, tiesinga1996spectroscopic}, the condensates's atomic number: $N = 5 \times 10^{6}$, and the reduced Planck constant: $\hbar = 1.05457 \times 10^{-34} J \cdot s$. Set the spacial unit as: $\beta = 100 \mu m =10^{-4}m $, the temporal unit as: $\alpha = \frac{2 M \beta^2}{\hbar} = 7.24 s $, the wave function's unit as: $\gamma = \sqrt{\frac{N}{\beta^3}} = 2.236 \times 10^{9} m^{-\frac{3}{2}}$, then the interaction parameter becomes: $\eta = \frac{8 \pi N a_s }{\beta} = 3.456 \times 10^{3}$. For the initial state, the distribution is supposed to satisfy the Gauss-type with the width parameter $\sigma = 1$ and the separation parameter $\delta_d = 2$ (under the spacial unit $\beta=10^{-4}m$), the initial momentum is related to the initial kinetic energy $E_k$ that: $p_{ini} = \sqrt{2 M E_k}$, which is $E_k = 0.5 n K \times k_B = 0.5 \times 10^{-9} K \times k_B$ in Ref. \cite{andrews1997observation} (where $k_B=1.38 \times 10^{-23} J/K$ is the Boltzmann constant), then the initial momentum parameter becomes: $\frac{\beta p_{ini}}{\hbar}= \frac{\beta \sqrt{2 M\times 0.5 \times 10^{-9} K \times k_B}}{\hbar} = 21.771$. Numerically solve the GP equation (\ref{eq-gpe-free-simple}) with the initial state in Eq. (\ref{eq-initial-separated}) and the parameters above, the results can be displayed as follows. 

\begin{figure}[ ]
	\centering
	\includegraphics[width = 0.45 \textwidth ]{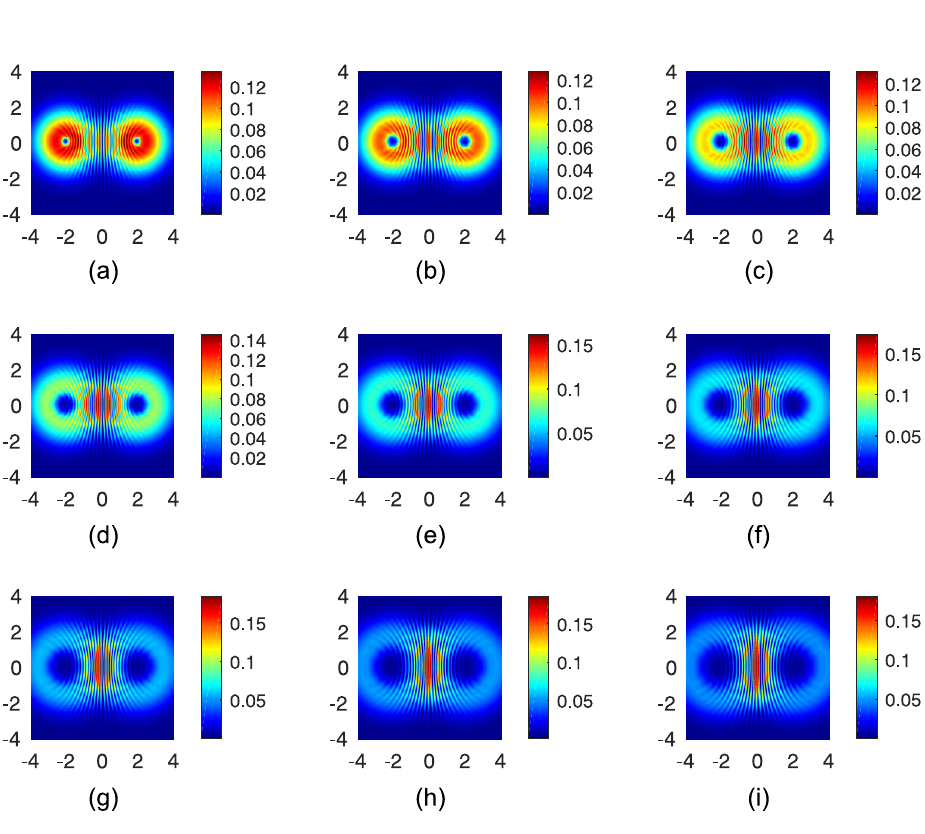}
	\caption{The interference of two Bose-Einstein condensates in two-dimensional space with the experimental parameters in Ref. \cite{andrews1997observation}. The spacial unit is set as: $\beta=10^{-4}m$, then the temporal unit becomes: $\alpha = 7.24 s$, and the other parameters becomes: $\sigma = 1$, $\delta_d = 2$, $\eta=3.456 \times 10^{3} $, $\frac{\beta p_{ini}}{\hbar} = 21.771 $. The evolution of $|\phi|^2$ has been shown from $\tau = 0.00221$ to $\tau = 0.1989$ (\textit{i.e.} time-of-flight from $16ms$ to $144ms$, in sub-figures from (a) to (i)), the temporal intervals between the neighbouring sub-figures are $0.00221$ ($16ms$). }
	\label{fig-experiment-density-tof*4}
\end{figure} 

In Fig. \ref{fig-experiment-density-tof*4}, the temporal evolution of the interference pattern between the two condensates with the experimental parameters in Ref. \cite{andrews1997observation} has been presented. The interference pattern will appear where the wave functions of the two condensates overlap. As time increases, the overlapping area of the wave functions of the two condensates expands, and the area where interference fringes appear also expands. 

\begin{figure}[ ]
	\centering
	\includegraphics[width = 0.45 \textwidth ]{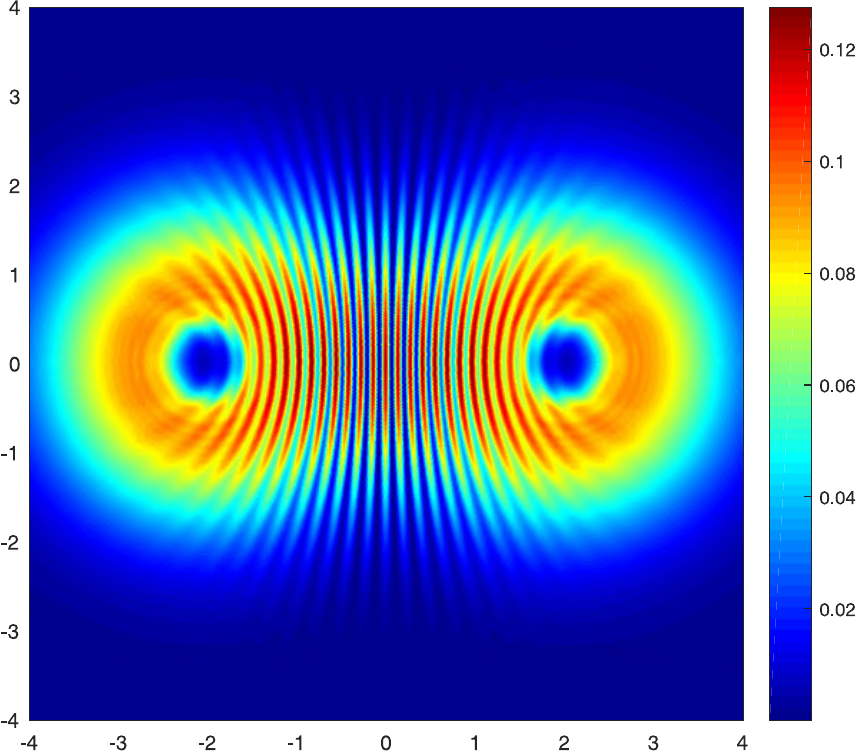}
	\caption{The interference pattern of the condensates from the solution of GP equation with the experimental data in Ref. \cite{andrews1997observation}. The pattern here is the density distribution of the condensates after the initial state has evolved for $40ms$ (dimensionless time variable: $\tau = 40ms / \alpha = 5.525 \times 10^{-3} $), corresponding to the $40 ms$ time-of-flight in Ref. \cite{andrews1997observation}. The other parameters are set as: $\eta=3.456 \times 10^{3}$, $\sigma = 1$, $\delta_d = 2$, $\frac{\beta p_{ini}}{\hbar}= 21.771$. The average spacing of the fringes here is about $14.81 \mu m$, which is very close to the experimentally observed fringe spacing $15 \mu m$ in Ref. \cite{andrews1997observation}.  }
	\label{fig-experiment-40ms}
\end{figure}

According to the case in Ref. \cite{andrews1997observation}, when the time-of-flight is $40ms$, the dimensionless time variable is: $\tau = 40ms / \alpha = 5.525 \times 10^{-3}$, the interference pattern from the numerical solution of the GP equation is presented in Fig. \ref{fig-experiment-40ms}. The interference fringes here are not exactly the same as the experimental results in Ref. \cite{andrews1997observation}, because the ideal two-dimensional case here is different from the quasi-two-dimensional case in experiment. Furthermore, the average spacing of the interference fringes here is about $14.81 \mu m$, that is close to the fringe spacing $15 \mu m$ in Ref. \cite{andrews1997observation}.

Consider the case in Ref. \cite{space-05-muntinga2013interferometry}, the experimental data here are: particle number: $N = 10^4$, rubidium-87 ($^{87}\mathrm{Rb}$) atomic mass: $M = 87 \times 1.66 \times 10^{-27}kg = 1.44 \times 10^{-25} kg$, the $s$-wave scattering length: $a_s = 95.44 a_0 = 5.05 \times 10^{-9}m$ (where $a_0 =5.29 \times 10^{-11} m$ is the Bohr radius) \cite{Rb-87-egorov2013measurement}. Set the spacial unit as: $\beta=10 \mu m$, the temporal unit as: $\alpha = \frac{2 M \beta^2}{\hbar} = 0.2731 s $, the wave function's unit as: $\gamma = \sqrt{\frac{N}{\beta^3}} = 3.162 \times 10^{9} m^{-\frac{3}{2}} $, then the interaction parameter becomes: $\eta = \frac{8 \pi N a_s }{\beta} = 126.92$. For the initial state, the effective temperature is about $T = 1 n K$ in Ref. \cite{space-05-muntinga2013interferometry}, then the related kinetic energy is: $E_k = 1 n K \times k_B = 1.38 \times 10^{-32} J$, and the initial momentum is: $p_{ini} = \sqrt{2 M E_k} = 6.3 \times 10^{-29} kg \cdot m/s$, so the initial momentum parameter becomes: $\frac{\beta p_{ini}}{\hbar}= 5.978 $. Let this initial states satisfy the seperated Gaussian distributions with the parameters: $\sigma = 1$ and $\delta_d = 2$, and numerical solve the  GP equation (\ref{eq-gpe-free-simple}), the results can be presented as follows. 

\begin{figure}[ ]
	\centering
	\includegraphics[width = 0.45 \textwidth ]{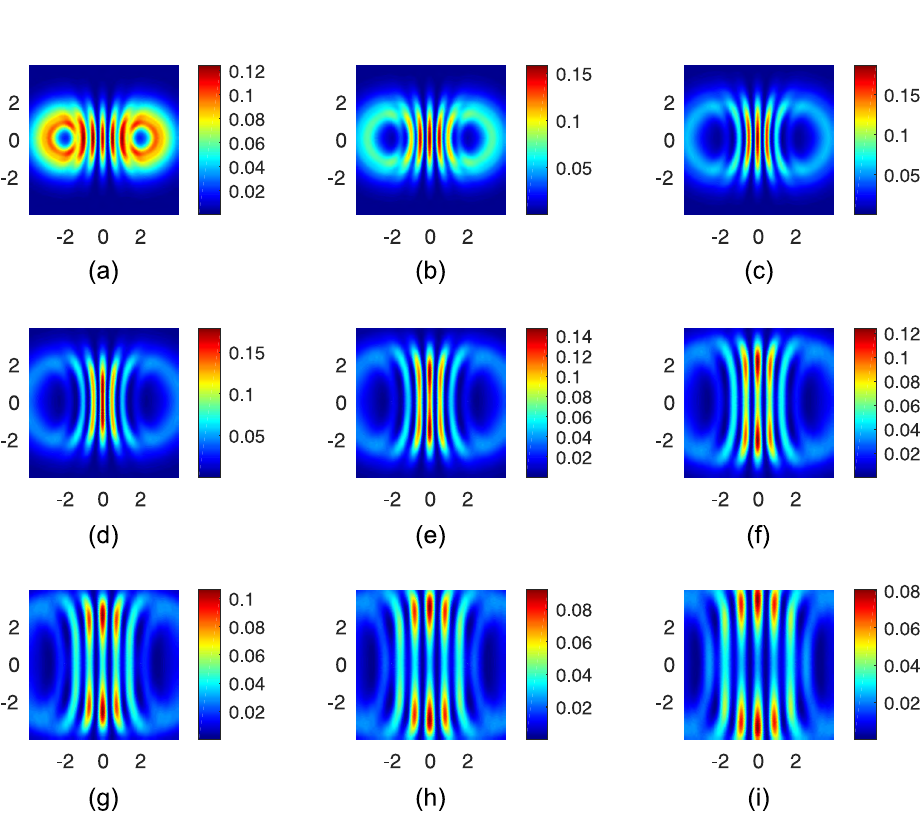}
	\caption{The interference of two Bose-Einstein condensates in two-dimensional space with the experimental parameters in Ref. \cite{space-05-muntinga2013interferometry}. The spacial unit is set as: $\beta=10^{-5}m$, then the temporal unit becomes: $\alpha = 0.2731 s$, and the other parameters becomes: $\sigma = 1$, $\delta_d = 2$, $\eta = 126.92 $, $\frac{\beta p_{ini}}{\hbar} = 5.978 $. The evolution of $|\phi|^2$ has been shown from $\tau = 0.02156667 $ to $\tau = 0.1941$ (\textit{i.e.} time-of-flight from $5.89ms$ to $53ms$, in sub-figures from (a) to (i)), the temporal intervals between the neighbouring sub-figures are $0.02156667$ ($5.89ms$).}
	\label{fig-experiment-2}
\end{figure}

In Fig. \ref{fig-experiment-2}, the temporal evolution of the interference pattern of two condensates with the experimental parameters in Ref. \cite{space-05-muntinga2013interferometry} has been presented. The interference fringes appear at where the wave functions of the condensates overlap. As time increases, the overlapping region of the wave functions gradually expands, and the spacing of the interference fringes increases slightly. 

\begin{figure}[ ]
	\centering
	\includegraphics[width = 0.45 \textwidth ]{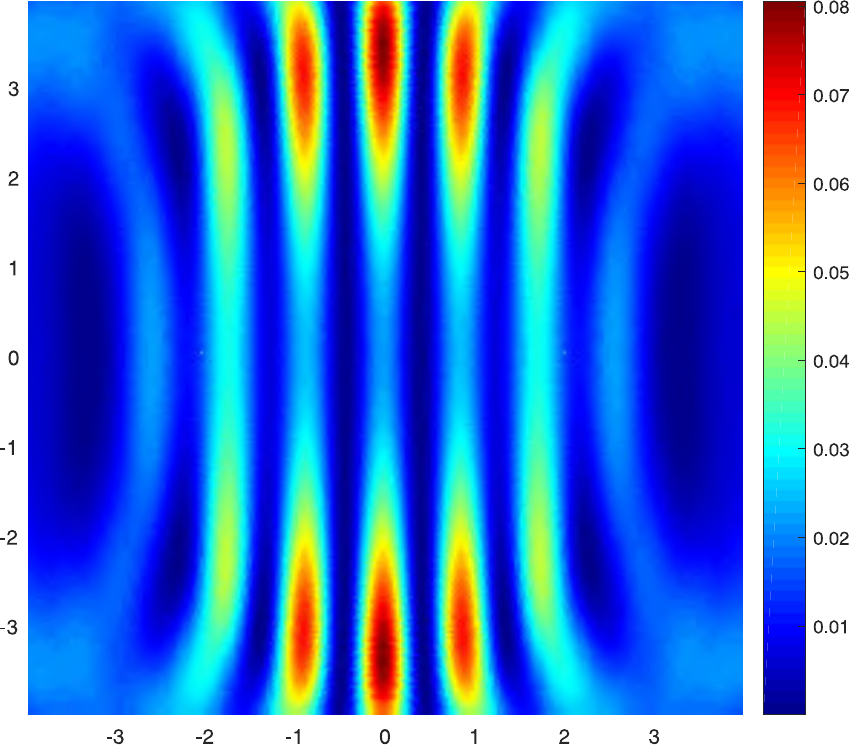}
	\caption{The interference pattern of the condensates from the solution of GP equation with the experimental data in Ref. \cite{space-05-muntinga2013interferometry}. The pattern here is the density distribution of the condensates after the initial state has evolved for $53ms$ (dimensionless time variable: $\tau = 53ms / \alpha = 0.1941 $), corresponding to the $53 ms$ time of expansion in Ref. \cite{space-05-muntinga2013interferometry}. The other parameters are set as: $\eta=126.92 $, $\sigma = 1$, $\delta_d = 2$, $\frac{\beta p_{ini}}{\hbar}= 5.978$. 
	}
	\label{fig-experiment-2-53ms}
\end{figure}

The interference pattern at $t=53ms$ ($\tau=0.1941$) with the experimental parameters in Ref. \cite{space-05-muntinga2013interferometry} has been presented in Fig. \ref{fig-experiment-2-53ms}. The average spacing of the interference fringes is about $10 \mu m$ in this figure, which is much less than the $75 \mu m \sim 107 \mu m$ in Ref. \cite{space-05-muntinga2013interferometry}. The deviation is supposed to come from the atomic number $N$, if there is a loss of the number of atoms in the experiment, the interaction strength cannot reach the theoretically estimated value, and the average spacing of the interference fringes will become larger. 

These results examine the effectiveness of GP equation for describing the dynamics of BECs, also supports the view of matter wave (wave-particle duality) which is one of the most fundamental concepts of quantum mechanics. We know that there are many unavoidable errors in various experiments, the results here are not completely consistent, and the relevant results need to be further confirmed by related experiments in the micro-gravity environment.

\section{Initial wave functions with different initial phase} 

Consider two condensates with an initial phase difference $\theta$, the initial wave function becomes: 
\begin{equation}
\label{eq-initial-phase difference}
	\begin{split}
		\phi_0 &= \phi_{ini}(\xi_1 + \delta_d,\; \xi_2) + \phi_{ini}(\xi_1 - \delta_d,\; \xi_2) \times e^{i \theta} \\
		&= \sqrt{\left(\frac{1}{\sqrt{2 \pi \sigma^2}} \right)^2 e^{-\frac{(\xi_1 + \delta_d)^2 + \xi_2^2}{2 \sigma^2} } } \times e^{\frac{i \beta}{\hbar} p_{ini} \sqrt{(\xi_1 + \delta_d)^2 + \xi_2^2}} \\
		&+ \sqrt{\left(\frac{1}{\sqrt{2 \pi \sigma^2}} \right)^2 e^{-\frac{(\xi_1 - \delta_d)^2 + \xi_2^2}{2 \sigma^2} } } \times e^{\frac{i \beta}{\hbar} p_{ini} \sqrt{(\xi_1 - \delta_d)^2 + \xi_2^2} + i\theta},
	\end{split}
\end{equation}
and solve GP equation (\ref{eq-gpe-free-simple}) with this initial state, the results can be shown as follows. 

\begin{figure}[ ]
	\centering
	\includegraphics[width = 0.45 \textwidth ]{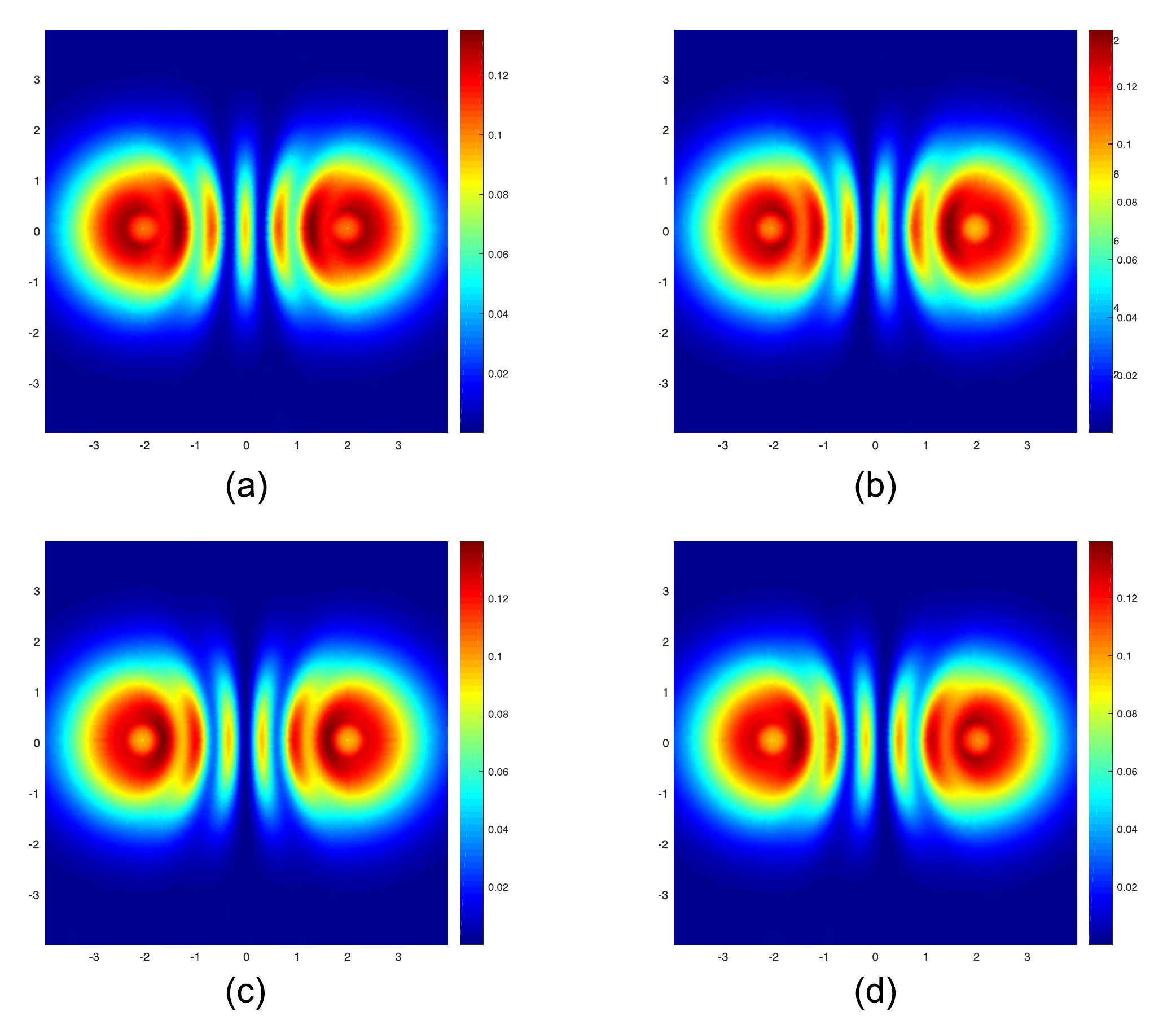}
	\caption{Interference pattern of two condensates with initial phase difference. Here, the common parameters are set as: $\tau=0.006$, $\sigma = 1$, $\delta_d = 2$, $\eta=3000$, $\frac{\beta p_{ini}}{\hbar}= 4$. The initial phase difference parameters are: $\theta=0$ in (a), $\theta= \pi /2$ in (b), $\theta=\pi$ in (c), $\theta= 3 \pi / 2$ in (d). }
	\label{fig-compare-phase difference}
\end{figure}

In Fig. \ref{fig-compare-phase difference}, the interference patterns of two condensates with different initial phase difference have been presented. The phase difference may cause the shift of the interference fringes. When $\theta=\pi/2$, the interference pattern is still symmetrical about the line $\xi_1=0$, which is shown in (c) of Fig. \ref{fig-compare-phase difference}. Let $\theta \rightarrow 2\pi - \theta$, the interference pattern will becomes the mirror image of line $\xi_1=0$, which is shown in (b) and (d) of Fig. \ref{fig-compare-phase difference}. It should be noted that interference of these two condensates here is only influenced by the relative phase between the two condensates. 

The initial phase of the wave function is also related to the initial momentum, let the two condensates have different initial momentum, set the additional phase difference as $\theta=0$, then the initial wave function becomes:
\begin{equation}
\label{eq-initial-different momentum}
	\begin{split}
		\phi_0 &= \phi_{ini}(\xi_1 + \delta_d,\; \xi_2, p_{ini, L}) + \phi_{ini}(\xi_1 - \delta_d,\; \xi_2, p_{ini, R}) \\
		&= \sqrt{\left(\frac{1}{\sqrt{2 \pi \sigma^2}} \right)^2 e^{-\frac{(\xi_1 + \delta_d)^2 + \xi_2^2}{2 \sigma^2} } } \times e^{\frac{i \beta}{\hbar} p_{ini, L} \sqrt{(\xi_1 + \delta_d)^2 + \xi_2^2}} \\
		&+ \sqrt{\left(\frac{1}{\sqrt{2 \pi \sigma^2}} \right)^2 e^{-\frac{(\xi_1 - \delta_d)^2 + \xi_2^2}{2 \sigma^2} } } \times e^{\frac{i \beta}{\hbar} p_{ini, R} \sqrt{(\xi_1 - \delta_d)^2 + \xi_2^2} },
	\end{split}
\end{equation}
where the $p_{ini, L}$ and the $p_{ini, R}$ are the initial momentums of the left condensate and the right condensate. Solve the GP equation (\ref{eq-gpe-free-simple}) with this initial state, the results can be presented as follows. 

\begin{figure}[ ]
	\centering
	\includegraphics[width = 0.45 \textwidth ]{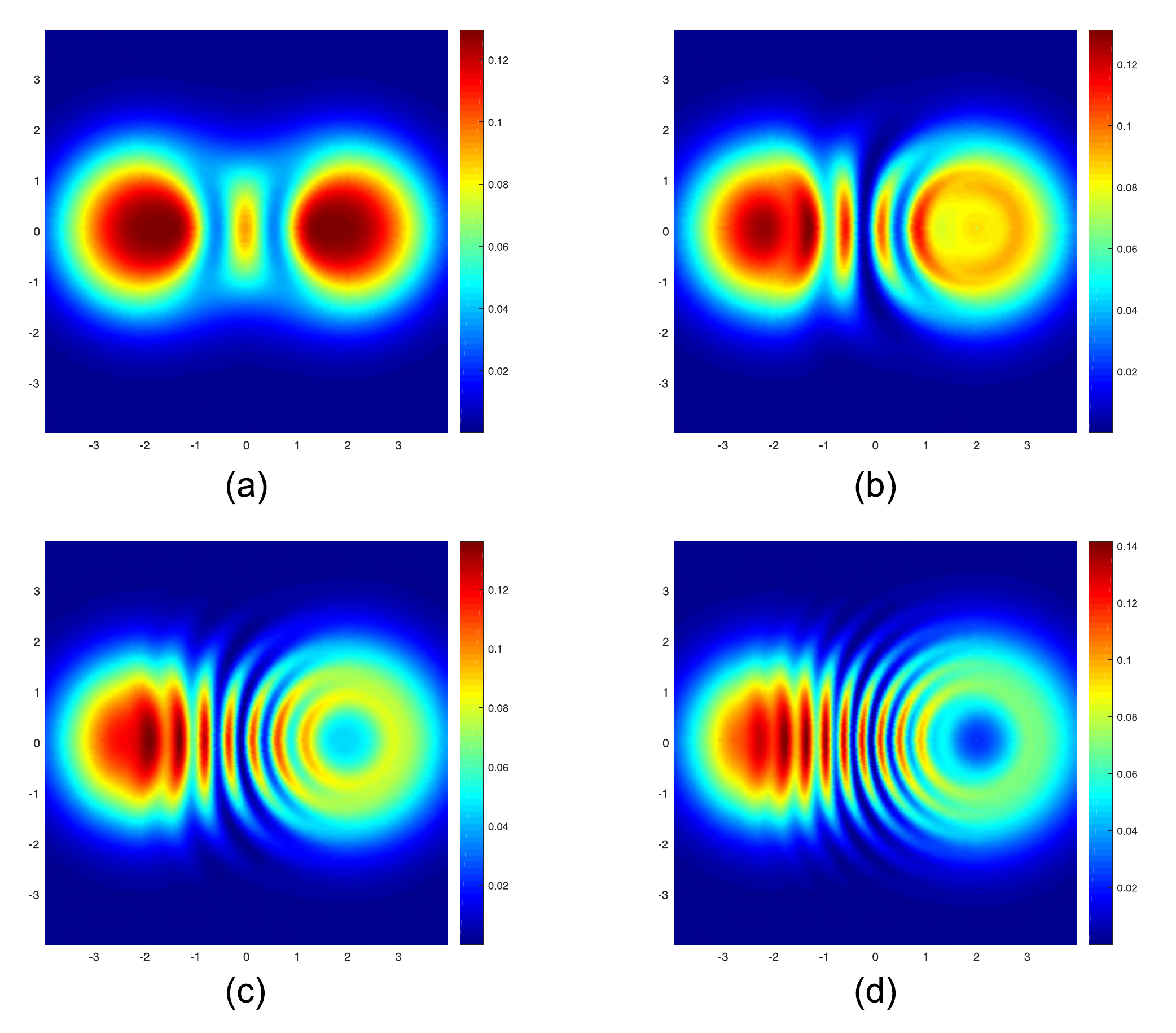}
	\caption{The interference pattern of the condensates with different initial momentum. The common parameters of the sub-figures are set as: $\tau=0.016$, $\sigma=1$, $\delta_d=2$, $\eta=3000$. The initial momentum parameter of the left condensate is set as: $\frac{\beta p_{ini, L}}{\hbar}= 0$, while those of the right condensate are: $\frac{\beta p_{ini, R}}{\hbar}= 0$ in (a), $\frac{\beta p_{ini, R}}{\hbar}= 4$ in (b), $\frac{\beta p_{ini, R}}{\hbar}= 8$ in (c), $\frac{\beta p_{ini, R}}{\hbar}= 12$ in (d).  }
	\label{fig-compare-kR, kL=0}
\end{figure}

\begin{figure}[ ]
	\centering
	\includegraphics[width = 0.45 \textwidth ]{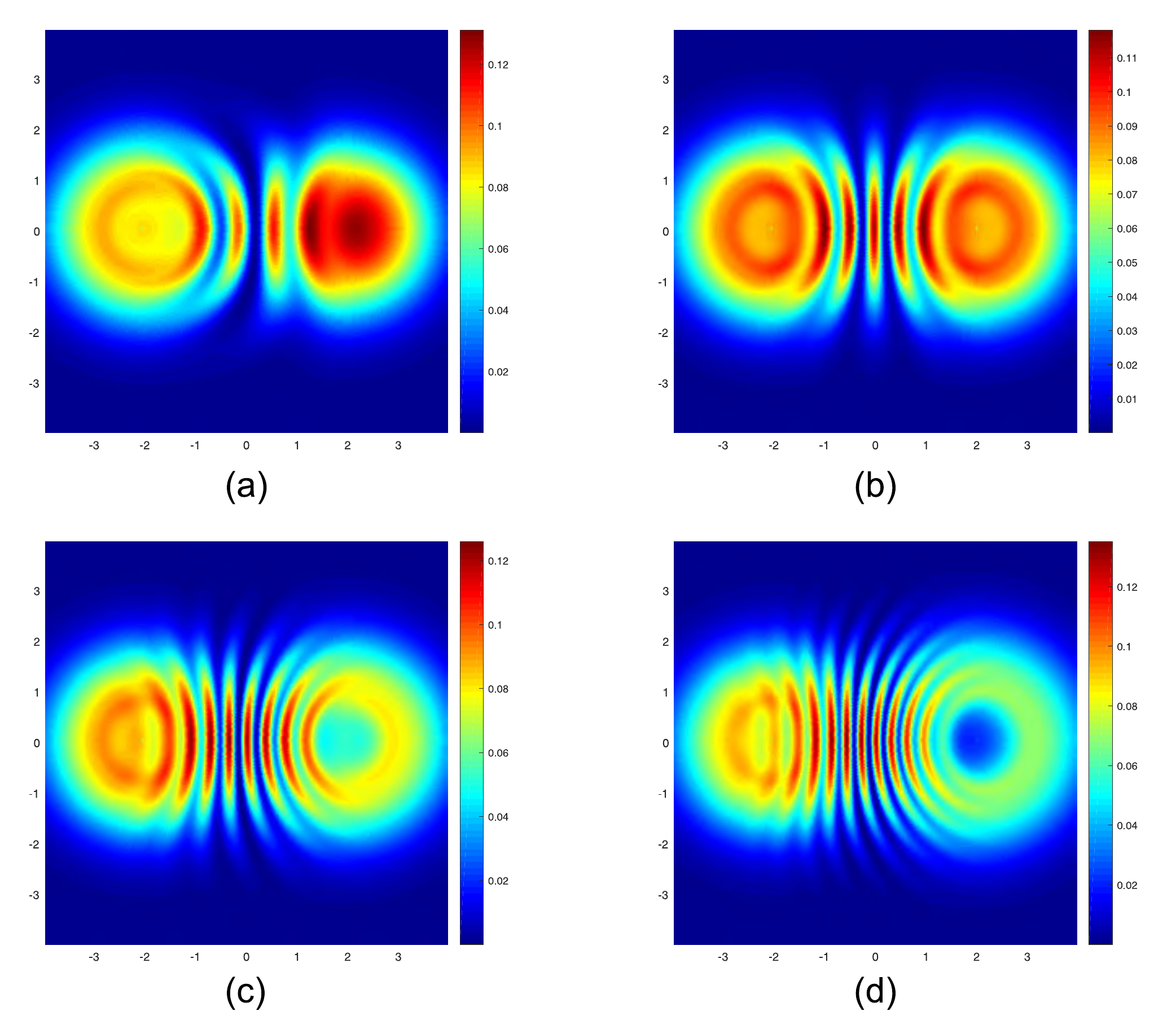}
	\caption{The interference pattern of the condensates with different initial momentum. The common parameters of the sub-figures are set as: $\tau=0.016$, $\sigma=1$, $\delta_d=2$, $\eta=3000$. The initial momentum parameter of the left condensate is set as: $\frac{\beta p_{ini, L}}{\hbar}= 4$, while those of the right condensate are: $\frac{\beta p_{ini, R}}{\hbar}= 0$ in (a), $\frac{\beta p_{ini, R}}{\hbar}= 4$ in (b), $\frac{\beta p_{ini, R}}{\hbar}= 8$ in (c), $\frac{\beta p_{ini, R}}{\hbar}= 12$ in (d).  }
	\label{fig-compare-kR, kL=4}
\end{figure}

In Figs. \ref{fig-compare-kR, kL=0} and \ref{fig-compare-kR, kL=4}, the interference patterns of the condensates with different initial momentum have been presented. In the case that one of the condensates has zero initial momentum, the interference fringes will also appear as the other condensate expand its region to overlap each other, which is shown in Fig. \ref{fig-compare-kR, kL=0}. If the initial momentum of the left condensate is not zero, as the right condensate's initial momentum increases, the right one tends to suppress the left one. At the some time, the interference fringes becomes denser, which is shown in Fig. \ref{fig-compare-kR, kL=4}. By the way, the condensate with bigger initial momentum may expand faster, as the initial momentum is related to the initial moving speed of the particles. According to the de Broglie's wavelength formula: $\lambda = h/p$ (where $h$ is the Planck's constant, $p$ is the momentum), the wavelength of the macroscopic wave function will be decreased as the momentum increases, which can be characterised by the decreased spacing of the interference fringes. These results are consistent with our physical intuition based on the classical and quantum mechanics about the matter wave and the related quantum state.

\section{Summary}

We have discussed the dynamical interference of two Bose-Einstein condensates in micro-gravity environment by numerically solving the Gross-Pitaeviskii equation with the initial state which satisfies the Gaussian distribution in two-dimensional space. Our results show that the increased interaction strength between the particles may accelerate the particles moving and make the interference fringes denser. The initial momentum of the particles is related to the number of the interference fringes. The condensate with zero initial momentum and energy may still have dynamical evolution, due to the macroscopic effective momentum and energy from the initial wave function's Gaussian distribution. The cases of gravity have discussed and the results show that the gravitational potential may shift the density and phase distributions of the wave function, which suggests that the experiments \cite{space-01-liu2018orbit, space-02-lachmann2021ultracold, space-03-carollo2022observation, space-04-gaaloul2022space, space-05-muntinga2013interferometry} in the microgravity environment are of great significance to explore the true nature of nature. 
The results of the simulation with the experimental parameters are generally consistent with the experimental results. The initial phase difference of the condensates may shift the interference fringes, the different initial momentum may also influence the interference pattern through the phase. These results are consistent with our physical intuition about the dynamics of the macroscopic wave function of the condensates. 
Our results are consistent with previous studies, verifying the consistency of the GP equation with experimental results and physical intuition, and providing assistance for the study of cold atoms in a micro-gravity environment where the physical environment is more ideal than usual. 
We also point out the importance of the macroscopic effective momentum and energy due to the macroscopic distribution of the wave function for the interference and dynamical evolution of the wave functions, which needs to be noted in future studies.

\begin{acknowledgments}

Thanks to Prof. Xing-Dong Zhao, Prof. Xiao-Fei Zhang, Prof. Jiang-Min Zhang, Dr. Qiao Yang, Dr. Cheng-Xi Li, Dr. Wan-Zi Sun for helpful discussions.  

This work was supported by National Key R\&D Program of China under grants No. 2021YFA1400900, 2021YFA0718300, 2021YFA1402100, NSFC under grants Nos.  61835013, 12174461, 12234012.

\end{acknowledgments}

\bibliography{CITE.bib}

\end{document}